\documentclass[11pt,prl,twocolumn,english,aps,reprint,superscriptaddress,notitlepage,nobibnote,preprintnumbers,showpacs]{revtex4-2}

\usepackage{graphicx}
\usepackage{booktabs}
\usepackage{amsmath}
\usepackage{mathrsfs}
\usepackage{amssymb}
\usepackage[export]{adjustbox}
\usepackage{bm}
\usepackage{color}
\usepackage{dcolumn}
\usepackage{xcolor}
\usepackage{autobreak}
\usepackage{multirow}
\usepackage{subfig}
\usepackage[vcentermath]{youngtab}
\usepackage{pifont}
\usepackage{extarrows}
\usepackage{makecell}
\usepackage{ulem}
\usepackage{bbm}
\usepackage{diagbox}
\usepackage{slashed}
\usepackage{tikz}
\usepackage[compat=1.1.0]{tikz-feynhand}
\usetikzlibrary{snakes}
\usetikzlibrary{arrows}

\usepackage[utf8]{inputenc}
\usepackage{times}

\newcommand{\beq}{\begin {equation}}
\newcommand{\eeq}{\end   {equation}}
\newcommand{\bea}{\begin {eqnarray}}
\newcommand{\eea}{\end   {eqnarray}}
\newcommand{\baa}{\begin {array}   }
\newcommand{\eaa}{\end   {array}   }
\newcommand{\bit}{\begin {itemize} }
\newcommand{\eit}{\end   {itemize} }
\newcommand{\be }{\begin {equation}}
\newcommand{\ee }{\end   {equation}}

\begin{document}

\title{\boldmath Geometric Formulation of Covariant Spin Operator}

\author{Yi-Ning Wang}
\affiliation{
CAS Key Laboratory of Theoretical Physics, Institute of Theoretical Physics, \\ Chinese Academy of Sciences, Beijing 100190, P.\ R.\ China}
\affiliation{
School of Physical Sciences, University of Chinese Academy of Sciences, Beijing 100049, P.\ R.\ China}

\author{Jiang-Hao Yu}
\affiliation{
CAS Key Laboratory of Theoretical Physics, Institute of Theoretical Physics, \\ Chinese Academy of Sciences, Beijing 100190, P.\ R.\ China}
\affiliation{
School of Physical Sciences, University of Chinese Academy of Sciences, Beijing 100049, P.\ R.\ China}
\affiliation{
School of Fundamental Physics and Mathematical Sciences, Hangzhou Institute for Advanced
Study, UCAS, Hangzhou 310024, China}

\begin{abstract}

In relativistic quantum theory, Lorentz boosts induce momentum-dependent Wigner rotations, complicating the coupling of particle spins in arbitrary frames. We describe particle spin spaces as fibers of an associate bundle over the velocity manifold, and different particle's spins residing at distinct fibers. Analogy to general relativity, we introduce parallel transport to bring spins at different velocities into a common frame, where they can be coupled. Using spinor-helicity variables, we construct the parallel transport operator along geodesics and show how it is uniquely determined by little group covariance. This construction gives explicit form of the covariant spin and orbital operators, and yields a covariant 3-point amplitude basis with definite spin and orbital angular momentum. Higher-point amplitudes can be constructively obtained and applied to the covariant partial wave expansion for amplitudes, form factors, and correlation functions, etc.

\end{abstract}

\maketitle

\textbf{\textit{  Introduction.}} Spin is a fundamental, quantized property of an elementary particle. In quantum mechanics, spin transforms as intrinsic angular momentum under spatial rotations, but its behavior under boosts differs between nonrelativistic and relativistic theories. In nonrelativistic quantum theory, the spin operator, $S_i=J_i-(\mathbf x\times\mathbf p)_i$, is invariant under Galilean boosts, which therefore do not mix spin and orbital degrees of freedom. In relativistic quantum theory, by contrast, spin operators such as the Newton--Wigner spin operator~\cite{Newton:1949cq,Polyzou:2012ut} generally undergo momentum-dependent Wigner rotations under Lorentz boosts. Consequently, a Lorentz boost generally mixes spin and orbital degrees of freedom.

In multiparticle systems, coupling individual spins to form states of definite total spin is essential for describing scattering, decays, and bound states. In the nonrelativistic setting, Galilean invariance allows spins to be coupled using Clebsch--Gordan coefficients independently of the system's overall motion. In the relativistic setting, however, spin transformations follow induced representations of the Poincar\'e group and involve momentum-dependent Wigner rotations. Although spins can be coupled in the center-of-mass (COM) frame, extending the spin coupling to arbitrary frames while keeping its covariant form has remained a longstanding challenge.

We realize the relativistic spin is defined locally in the momentum space, treated as the bundle structure over the velocity manifold. Each velocity carries a spin space, represented by a fiber of an associated bundle. Spins at different velocities must be referred to a common frame before they can be coupled covariantly.  Analogy to parallel transport in general relativity, we define the transport to carry spins in different fibers into a common point along the geodesic path. This construction defines an additive covariant spin operator for a multiparticle system and a corresponding orbital operator. It also provides a geometric route from Wigner states to a covariant orbital--spin amplitude basis.

\medskip
\textbf{\textit{  Wigner state, spin operator and fiber bundle.}} In relativistic quantum theory a massive elementary particle is defined as a unitary irreducible representation of the Poincar\'e group, and Wigner's method of induced representations~\cite{Wigner:1939cj,Weinberg:1995mt} builds it from two ingredients. The first is the little group $H$, the Lorentz subgroup that stabilizes a chosen reference velocity $u_\star=(1,\mathbf 0)$: its unitary irreducible representations label the rest-frame states $|mu_\star,\sigma\rangle$. The second is the coset space $G/H \equiv \mathrm{SL}(2,\mathbb C)/\mathrm{SU}(2)$, whose representatives $L_u\in G$ label the velocities $u=L_uu_\star$ up to a rest-frame rotation $h(u_\star)\in H$, the momentum-dependent states being obtained by the boost action, $|p,\sigma\rangle=\hat U(L_u)|mu_\star,\sigma\rangle$ with $p=mu$. The set of all four-velocities $u$, that is, the coset space $G/H$, forms the velocity manifold
\begin{eqnarray}
M= \{u:u^2=1,\ u^0>0\} \simeq \mathrm{SL}(2,\mathbb C)/\mathrm{SU}(2).
\end{eqnarray}
Its Riemannian metric is $ds^2=-du\cdot du$ on tangent vectors, and it has constant curvature. A fixed-mass momentum shell is a rescaled copy of $M$; particles of different masses are therefore compared on the same normalized manifold, not on one fixed-mass shell.
The little group is the stabilizer of a chosen reference point $u_\star$ of $M$. At any other point $u=g u_\star$ of $M$, with $g\in G$, the corresponding stabilizer is the conjugate subgroup $gHg^{-1}$. Every $g$ admits the coset decomposition $g=L_u\mathsf h$ with $\mathsf h\in H$, and two choices of the boost representative $L_u$ differ by a rest-frame rotation $h(u)\in H$; that rotation is the freedom of choosing the spin axis at the velocity $u$, and geometrically it is a gauge transformation~\footnote{A Lorentz transformation $L\in\mathrm{SL}(2,\mathbb C)$ acts on a four-velocity through its Hermitian matrix representative: with $\mathsf u\equiv u_\mu\sigma^\mu$ one has $\mathsf u\mapsto L\,\mathsf u\,L^\dagger$, which is the double cover $\mathrm{SL}(2,\mathbb C)\to\mathrm{SO}^+(1,3)$. The velocity reached from $u_\star$ is written $u=Lu_\star$, so that $\mathsf u=L\,\mathsf u_\star L^\dagger$.}.

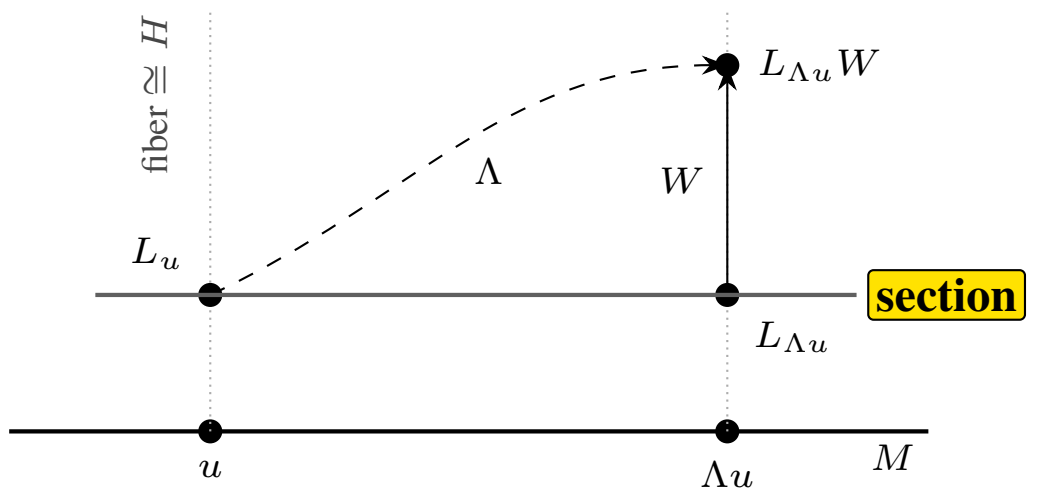
\begin{figure}[t]
\centering
\begin{tikzpicture}[>=Stealth,dot/.style={fill,circle,inner sep=1.6pt},font=\scriptsize,scale=0.95]
\draw[thick] (-0.4,0) -- (6.0,0);
\node[below left=-1pt] at (6.0,0) {$M$};
\node[dot,label=below:$u$] (u) at (1,0) {};
\node[dot,label=below:$\Lambda u$] (Lu) at (4.6,0) {};
\draw[gray!65,densely dotted] (1,0) -- (1,2.95);
\draw[gray!65,densely dotted] (4.6,0) -- (4.6,2.95);
\node[gray!55!black,rotate=90] at (0.62,2.35) {fiber $\cong H$};
\node[dot,label=above left:$L_u$] at (1,0.95) {};
\node[dot,label=below right:$L_{\Lambda u}$] at (4.6,0.95) {};
\node[dot,label=right:$L_{\Lambda u}W$] at (4.6,2.55) {};
\draw[->,dashed] (1,0.95) to[out=25,in=180] node[below right=-1pt] {$\Lambda$} (4.6,2.55);
\draw[->] (4.6,0.95) -- node[left=1pt] {$W$} (4.6,2.55);
\draw[thick,gray!75!black] (0.2,0.95) -- (5.5,0.95);
\node[draw,fill=yellow!85!orange,rounded corners=1pt,inner sep=1.5pt,right=2pt,font=\bfseries] at (5.5,0.95) {section};
\end{tikzpicture}
\caption{The coset decomposition on principal bundle. The left action of $\Lambda$ carries the frame $L_u$ into fiber over $\Lambda u$, and the Wigner rotation $W$ is the right multiplication with $L_{\Lambda u}$.}
\label{fig:PrincipleBundleSection}
\end{figure}

Lifting this coset picture to a bundle, $G$ forms a principal $H$-bundle over the coset base manifold $G/H$, with each velocity point $u$ equipped with a little-group $H$ fiber. A point of the bundle can be written $(u,\mathsf h)$; choosing the standard boosts $L_u$ and using the coset decomposition identifies two equivalent bundle structures~\cite{Dubrovin:1985,KobayashiNomizu:1969} $(u,\mathsf h)\cong(L_u,\mathsf h)=L_u\mathsf h$, which is a point $g\in G$.  A right multiplication $g\mapsto gh(u)$ with $h(u)\in H$ defines the right action: it changes the section, $L_u\to L_uh(u)$, and the fiber coordinate, $\mathsf h\to h(u)^{-1}\mathsf h\,h(u)$, since $g\,h(u)=L_u\mathsf h\,h(u)=\big(L_uh(u)\big)\big(h(u)^{-1}\mathsf h\,h(u)\big)$; $H$ is the gauge freedom that rotates the spin axis at fixed velocity and does not move $u$. Left action on $g$ is defined by a Lorentz  multiplication from the left, $g\mapsto\Lambda g=(\Lambda L_u,\mathsf{h})$. Using the coset decomposition $\Lambda L_u=L_{\Lambda u}W$ with the Wigner rotation $W(\Lambda,u)\equiv L_{\Lambda u}^{-1}\Lambda L_u$, one has $\Lambda g=(L_{\Lambda u},W\mathsf h)$: the left action moves the velocity, $u\mapsto\Lambda u$, and rotates the fiber by $W$ as shown in Fig.~\ref{fig:PrincipleBundleSection}.

The above tells that $g$ carries a left $G$-action and a right $H$-action, which act on the two indices of $g_\alpha{}^I$ separately; the group element is therefore conveniently labelled by the spin-spinor~\cite{Arkani-Hamed:2017jhn,Dittmaier:1998nn}. In the spinor-helicity formalism~\cite{Dixon:2013uaa,Elvang:2015rqa,Cheung:2017pzi,Badger:2023eqz} it reads
\begin{equation}\label{eq:spinorcoords}
\lambda_\alpha{}^I=\sqrt m\,g_\alpha{}^I,\quad
\tilde\lambda_{\dot\alpha I}=(\lambda_\alpha{}^I)^*,\quad
p_{\alpha\dot\alpha} =\lambda_\alpha{}^I\tilde\lambda_{\dot\alpha I},
\end{equation}
with $I$ the little-group index and $\alpha,\dot\alpha$ the Lorentz indices. Once we fix $g$ to be the standard boost $L_u$, the spinor transforms covariantly under the $G$-action as $\lambda_\alpha^I\mapsto\Lambda \lambda_\alpha^I(u)=\lambda_\alpha^J(\Lambda u)W_J^I(\Lambda,u)$, and under the $H$-action as $\lambda_\alpha^I\mapsto\lambda_\alpha^J h_J^I$.

The spin operator is the infinitesimal generator of the right action of $H$, that is, of the transformation that rotates the spin axis at fixed velocity. Using the coordinate $\lambda_\alpha^I=\sqrt{m}g_\alpha^I$ on the principal bundle, its differential realization in terms of vector field $\partial/\partial \lambda_\alpha^I$ is:
\begin{equation}\label{eq:Sprincipal}
S_I{}^J\equiv\lambda^{(J}_\alpha\frac{\partial}{\partial\lambda^{I)}_\alpha}+\text{c.c.},\qquad
S_\rho\equiv\tfrac12(\sigma_\rho)^I{}_J\,S^J{}_I,
\end{equation}
where parentheses denote symmetrization in $I,J$, and c.c. denotes the complex conjugation and $\rho=\pm1,0$ labels the spherical basis of Pauli matrices.

Once a section $L_u$ is chosen, the spin operator can also be written in terms of the Poincar\'e generators. With the Lorentz generators $M_{\alpha\beta}\equiv\lambda_{(\beta}^I\,\partial/\partial\lambda^{\alpha)I}$, the Pauli--Lubanski vector is $\mathcal W_{\alpha\dot\alpha}=p^\beta{}_{\dot\alpha}M_{\alpha\beta}-p^{\dot\beta}{}_\alpha\tilde M_{\dot\alpha\dot\beta}$, and the polarization vectors of the section are $\epsilon^{IJ}_{\alpha \dot\alpha}(u)\cong\frac{1}{m}\lambda^{(I}_{\alpha}\tilde\lambda^{J)}_{\dot \alpha}$; expanding Eq.~\eqref{eq:Sprincipal} in the spinor coordinates then gives
\begin{equation}\label{eq:Spoincare}
S_{IJ}(u)=\frac{1}{2m}\,\epsilon^{\alpha\dot\alpha}_{IJ}(u)\,\mathcal W_{\alpha\dot\alpha}, \quad 
S_{\rho}(u)=\frac{1}{m}\,\epsilon^\mu(u)_\rho\,\mathcal W_\mu.
\end{equation}
Taking $L_u$ to be the canonical boost $L_u=B_c(u)$ gives the Newton-Wigner spin operator $S_i(u)=\frac{1}{m}\Big(\mathcal W_i-\frac{\mathcal W^0p_i}{p^0+m}\Big)$.

The principal bundle $G$ has an associated bundle $E_s\equiv G\times_H V_s$ to describe massive spin-$s$ particles, with the spin space $V_s$ carrying spin-$s$ irreducible representation $\rho_s$ of $H=\mathrm{SU}(2)$. The fiber $(E_s)_u$ at velocity $u$ is then the representation space of the spin operator $S_{IJ}(u)$. A spin-$s$ particle state is built from Wigner states together with a wave function $\psi_\sigma(u)$ over $E_s$,
\begin{equation}\label{eq:Hilbertisomorphism}
|\Psi\rangle=\sum_{\sigma=-s}^{s}\int_M\tfrac{d^3\mathbf u}{u^0}\,\psi_\sigma(u)\,|mu,\sigma\rangle,
\end{equation}
with $d^3\mathbf u/u^0$ the Lorentz-invariant measure on $M$. These states span the Hilbert space $\mathcal H_s\cong L^2(M,\tfrac{d^3\mathbf u}{u^0})\otimes V_s$, that is, the space of square-integrable functions on $E_s$.

Since $E_s$ is associated to $G$ through $\rho_s$, the vertical vector fields of Eq.~\eqref{eq:Sprincipal} act on the fiber $V_s$ through the differential of $\rho_s$, yielding the fiber-wise operators $S_\rho\,\psi_\sigma (u)=\left(\mathcal J_\rho\right)_{\sigma\sigma' }\,\psi_{\sigma'}(u)$, where $\mathcal J_\rho$ are the spin-$s$ angular-momentum matrices in the spherical basis.
The unitary isomorphism Eq.~\eqref{eq:Hilbertisomorphism}, which identifies the component $\psi_\sigma(u)$ of the wave function with the coefficient of $|mu,\sigma\rangle$, gives the quantum operator action $S_\rho|mu,\sigma\rangle=\sum_{\sigma'}|mu,\sigma'\rangle\big(\mathcal J_\rho\big)_{\sigma'\sigma}$.
%\begin{equation}\label{eq:Squantum}
%\end{equation}

From Eq.~\eqref{eq:Sprincipal} and Eq.~\eqref{eq:Spoincare}, the spin operator is defined relative to the chosen $L_u$, and therefore depends on the choice of spin axes. This is what obstructs a direct coupling of two particle spins. Under $L_u\to L_uh(u)$, the spin operator rotates according to $S_\rho\mapsto \rho_1(h^{-1})_\rho{}^\sigma S_\sigma$, with $\rho_1$ the adjoint (spin-one) representation: it is gauge covariant rather than gauge invariant, carrying one adjoint little-group index. A sum of two such operators is therefore not covariant: under a change of frame, $S_\rho(u)+S_\rho(v)$ becomes
\begin{equation}\label{eq:sumnotcovariant}
\rho_1\big(h(u)^{-1}\big)_\rho{}^\sigma\,S_\sigma(u)
+\rho_1\big(h(v)^{-1}\big)_\rho{}^\sigma\,S_\sigma(v),
\end{equation}
because the two frames can be redefined independently, by $h(u)$ and by $h(v)$; the sum acquires two different rotations and is no longer of the same form. No natural unified choice of the two frames exists that would repair this, so a direct coupling of the two spins is impossible; they can be coupled only after being referred to a common frame.

\medskip
\textbf{\textit{  Spin transport along geodesic.}}
For the spin sum $ S_\rho(u)+S_\rho(v)$ to transform covariantly under the gauge transformation, $S^{\rm tot}\to\rho(h^{-1})S^{\rm tot}$ with $h\in H$, we introduce a transport $T(v,u)\in H$ that refers the spin at $u$ to the frame at $v$. The transported sum $S^{\rm tot}_\rho=\rho_1\big(T(v,u)\big)_\rho{}^\sigma S_\sigma(u)+S_\rho(v)$ is a little-group tensor at $v$ only if the transport compensates the two independent redefinitions of the frames, that is, only if
\begin{equation}\label{eq:Tgauge}
T(v,u)\ \longrightarrow\ h(v)^{-1}\,T(v,u)\,h(u)
\end{equation}
under $L_u\to L_uh(u)$ and $L_v\to L_vh(v)$: with this gauge law the two little-group factors in the transported sum combine into the single rotation $\rho_1(h(v)^{-1})$ of an operator at $v$, and the sum keeps its form.

%\textit{Transport.} 

To construct $T(v,u)$, an element of $H$ obeying the gauge law of Eq.~\eqref{eq:Tgauge}, we need a connection on the principal bundle. There a single element of $H$ encodes the transport for every representation, the action on $E_s$ then following by applying $\rho_s$ fiberwise. The canonical connection is fixed by the Cartan decomposition $\mathfrak g=\mathfrak h\oplus\mathfrak m$, with $\mathfrak h=\mathfrak{su}(2)$ the little-group algebra: its form is the $\mathfrak h$-component of the Maurer--Cartan form~\cite{Coleman:1969sm,Callan:1969sn,KobayashiNomizu:1969,Eguchi:1980jx}: $\lambda^{-1}\mathrm d\lambda=\omega_h+\omega_m$, with the vertical component $\omega_h\in\mathfrak h$, while the horizontal component $  \omega_m\in\mathfrak m$ dropped. To compare the frames at two different velocities one integrates along a curve $\gamma$ in $M$ joining them, and the transport is the path-ordered exponential of $\omega_h$,
\begin{equation}\label{eq:generictransport}
T_\gamma=\mathcal P\exp\Big(-\int_\gamma\omega_h\Big)\in H,
\end{equation}
which belongs to the little group fiber, and thus the gauge law $T(v,u)\mapsto h(v)^{-1}T(v,u)h(u)$ under the right action $L_u\to L_uh(u)$ follows.

In addition to their behavior under changes of section, both $S_\rho$ and $T(v,u)$ must respect the left action of the Lorentz transformation. For the spin operator, Lorentz transformation induces the Wigner rotation $W(\Lambda,u)$ of the frame at $u$, and pushing this to $\mathcal H_s$ through the isomorphism Eq.~\eqref{eq:Hilbertisomorphism} gives
\begin{equation}\label{eq:Spintransform}
U(\Lambda)\,S_\rho(u)\,U(\Lambda)^{-1}
=\rho_1\big(W(\Lambda,u)^{-1}\big)^{\;\;\sigma}_{\rho}S_\sigma(\Lambda u)\,,
\end{equation} 
where the Wigner rotation $W(\Lambda, u)$ depends on $u$ and the chosen section. Therefore, the same $\Lambda$ gives different Wigner rotations at different velocities,
\begin{equation}\label{eq:littlegroups}
W(\Lambda,u)\neq W(\Lambda,v)\qquad(u\neq v),
\end{equation}
so the direct sum $S(u)+S(v)$ does not generally transform by a common rotation. On the other hand, the transported sum, written schematically as $S^{\rm tot} = T(v,u)S(u)+S(v)$, obeys
$U(\Lambda)\,S^{\rm tot}_\rho(v)\,U(\Lambda)^{-1}
=\rho_1\big(W(\Lambda,v)^{-1}\big)^{\;\;\sigma}_{\rho}S^{\rm tot}_\sigma(\Lambda v)$, 
only if the transport itself compensates the two rotations,
\begin{equation}\label{eq:symOfT}
T(\Lambda v,\Lambda u)=W(\Lambda,v)\,T(v,u)\,W(\Lambda,u)^{-1},
\end{equation}
so that the two little-group factors combine into the single rotation of an operator at $\Lambda v$.

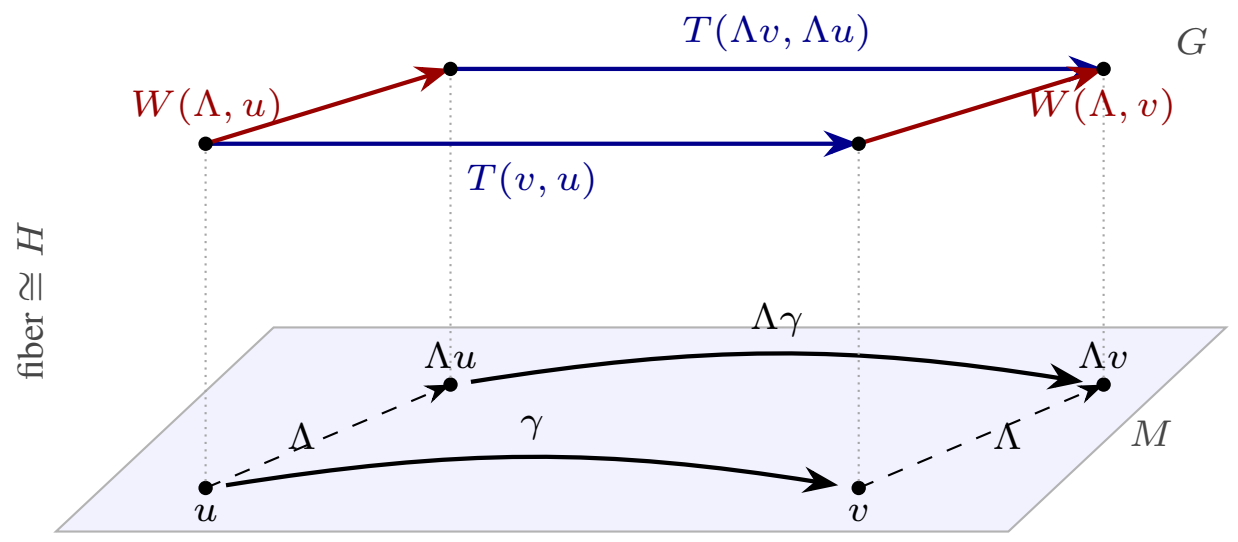
\begin{figure}[tb!]
\centering
\begin{tikzpicture}[>=Stealth,font=\scriptsize,scale=0.9]
\fill[blue!5] (0,0) -- (7.0,0) -- (8.6,1.5) -- (1.6,1.5) -- cycle;
\draw[gray!60] (0,0) -- (7.0,0) -- (8.6,1.5) -- (1.6,1.5) -- cycle;
\node[gray!55!black] at (8.05,0.72) {$M$};
\fill (1.1,0.32) circle (1.5pt) node[below] {$u$};
\fill (5.9,0.32) circle (1.5pt) node[below] {$v$};
\fill (2.9,1.08) circle (1.5pt) node[above] {$\Lambda u$};
\fill (7.7,1.08) circle (1.5pt) node[above] {$\Lambda v$};
\draw[->,thick] (1.25,0.34) to[out=9,in=171] node[above] {$\gamma$} (5.75,0.34);
\draw[->,thick] (3.05,1.10) to[out=9,in=171] node[above] {$\Lambda\gamma$} (7.55,1.10);
\draw[->,dashed] (1.1,0.32) -- (2.9,1.08) node[midway,left=-1pt] {$\Lambda$};
\draw[->,dashed] (5.9,0.32) -- (7.7,1.08) node[midway,right=-1pt] {$\Lambda$};
\draw[gray!70,densely dotted] (1.1,0.37) -- (1.1,2.85);
\draw[gray!70,densely dotted] (5.9,0.37) -- (5.9,2.85);
\draw[gray!70,densely dotted] (2.9,1.13) -- (2.9,3.4);
\draw[gray!70,densely dotted] (7.7,1.13) -- (7.7,3.4);
\node[gray!55!black,rotate=90] at (-0.18,1.7) {fiber $\cong H$};
\node[gray!55!black] at (8.35,3.6) {$G$};
\draw[->,thick,blue!55!black] (1.1,2.85) -- (5.9,2.85) node[midway,below] {$T(v,u)$};
\draw[->,thick,blue!55!black] (2.9,3.4) -- (7.7,3.4) node[midway,above] {$T(\Lambda v,\Lambda u)$};
\draw[->,thick,red!60!black] (1.1,2.85) -- (2.9,3.4) node[midway,left=5pt] {$W(\Lambda,u)$};
\draw[->,thick,red!60!black] (5.9,2.85) -- (7.7,3.4) node[midway,right=5pt] {$W(\Lambda,v)$};
\fill (1.1,2.85) circle (1.5pt); \fill (5.9,2.85) circle (1.5pt);
\fill (2.9,3.4) circle (1.5pt); \fill (7.7,3.4) circle (1.5pt);
\end{tikzpicture}
\caption{The principal bundle $G$ over the velocity manifold $M$: $M$ is the shaded patch, the dotted lines are the fibers
$H$. The path $\gamma$ joins the velocities $u$ and $v$, and a Lorentz
transformation carries the whole configuration into $\Lambda\gamma$, joining
$\Lambda u$ and $\Lambda v$. The lifts of the two paths induce the transports
$T$ (blue), while the frames at $u$ and $\Lambda u$, and at $v$ and
$\Lambda v$, are related by the Wigner rotations $W$ (red). In both diagrams the arrows commute: the two ways from
$u$ to $\Lambda v$ compose to the same map, Eq.~\eqref{eq:pathequivariance} on
the base and Eq.~\eqref{eq:symOfT} on the fibers.}
\label{fig:bundle}
\end{figure}

The transport $T_\gamma$ itself depends on the curve $\gamma$ chosen for it. The transport relation in Eq.~\eqref{eq:symOfT} would determine the $\gamma$ path to be geodesic. Let us start from each side of the Eq.~\eqref{eq:symOfT}, there are two unrelated curves, $\gamma(v,u)$ from $u$ to $v$ and $\gamma(\Lambda v,\Lambda u)$ from $\Lambda u$ to $\Lambda v$, and nothing connects them. To make it hold we must therefore require the curves themselves to be related, point by point (Fig.~\ref{fig:bundle})
\begin{equation}\label{eq:pathequivariance}
\gamma(\Lambda v,\Lambda u)=\Lambda\gamma(v,u).
\end{equation}
This condition fixes the curve uniquely: it is the geodesic path
\begin{equation}\label{eq:geodesic}
\gamma(s)=\frac{(1-s)u+sv}{\sqrt{(1-s)^2+s^2+2s(1-s)\,u\cdot v}},\qquad 0\leq s\leq1,
\end{equation}
which is built from the two velocities alone and therefore satisfies Eq.~\eqref{eq:pathequivariance} by construction; its parameter $s$ is in general not affine. The corresponding geodesic boost is $B(v,u)=\frac{\mathbf 1+\slashed v\,\overline{\slashed u}}{\sqrt{2(1+u\cdot v)}}$, and the transport along the geodesic could be evaluated although quite complicated. In this work, however, we would adapt more elegant way to construct the covariant spinor form of the transport in the principal bundle from symmetry argument, irrelevant of the explicit form of the geodesic and Lorentz boost, provided in supplementary material.

\textbf{\textit{Transport and on-shell amplitudes in spinor formalism.}}  The transport $T$ is constrained by Eq.~\eqref{eq:symOfT}, which indicates that $T$ carries one little-group index at $u$ and one at $v$, so that $T(v,u)$ can be written as a function of the spinors $\lambda(u)$ and $\lambda(v)$. For constituents of spin $\tfrac12$ the transport is a $2\times 2$ matrix $T^I{}_K$; like the form of scattering amplitudes below, it is a little-group tensor, carrying one index at $u$ and one at $v$. 

Just like the three-point amplitudes are uniquely determined by the Poincare symmetry~\cite{Arkani-Hamed:2017jhn,Benincasa:2007xk,Witten:2003nn}, we note two requirements determine the amplitude form of the transport completely, although only two momenta $u, v$ are involved in $T$. (i) \textit{Symmetry}: a Lorentz-invariant object carrying one index at each leg can only be built from the two spinors $\lambda(u)$ and $\lambda(v)$, and the only invariant contractions with one index at each leg are $\langle u_Kv^I\rangle$ and $[u_Kv^I]$, a mixed contraction $\lambda\tilde\lambda$ carries a Lorentz index and is excluded, while every other contraction is a scalar $u\cdot v$ that belongs to the coefficient,
\begin{equation}\label{eq:Ttwostructures}
T^I{}_K(v,u)=a\,\langle u_Kv^I\rangle+b\,[u_Kv^I].
\end{equation}
(ii) \textit{Kinematics}: choosing the canonical section $L_u=B_c(u)$ and $L_v=B_c(v)$, we note in the COM frame the spin coupling is trivial, $T^I_K(v,u_\star)=\delta^I_K$, which fixes $b=-a$ and the normalization,
\begin{equation}\label{eq:tau}
\tau^I_{iK}\equiv[v^Iu_K]-\langle v^Iu_K\rangle,\,
T^I_{iK}=\frac{\tau^I_{iK}}{\sqrt{2m_um_v\,(u\cdot v+1)}} .
\end{equation}
The normalization $m_um_v$ comes from the $\sqrt m$ in Eq.~\eqref{eq:spinorcoords}: for the physical momenta $m_uu$ and $m_vv$, the $\tau$ above carries dimension $[\tau]=[\text{mass}]$. Eq.~\eqref{eq:tau} is the unique matrix carrying one index at $v$ and one at $u$ and reducing to $\delta^I_K$ in the COM frame, and the horizontal lift along a geodesic has precisely these properties by construction. The symmetric construction above is consistent with the earlier geometric construction using $\omega_h$ as in Eq.~\eqref{eq:symOfT}. For spin $s_i$ it acts on the symmetric index set $\{I\}=(I_1\cdots I_{2s_i})$ as the symmetrized product
\begin{equation}\label{eq:taupower}
\rho_s(T)^{\{I\}}_{\{K\}}=T^{(I_1}_{(K_1}\cdots T^{I_{2s_i})}_{K_{2s_i})},
\end{equation}
so that a single spin-$\tfrac12$ object generates every spin.

\medskip

 With $T$ at hand the spins of several particles can be brought into a single frame. Transporting each little-group index separately gives the transported spin operator,
\begin{equation}\label{eq:parallelTransport}
\mathcal{S}_{K_1K_2}(u)=T_{K_1}{}^{I_1}(U,u)T_{K_2}{}^{I_2}(U,u)\,S_{I_1I_2}(u),
\end{equation}
the general case being the symmetric product $\rho_s(T)^{\{I\}}_{\{K\}}$ of Eq.~\eqref{eq:taupower}, the little-group indices at $u_i$ being contracted in favour of those at $U$, the transported spin operators all carry the same indices and can be added,
\begin{equation}\label{eq:spinoperatorcoupling}
\mathcal{S}_{K_1K_2}(U)=\mathcal{S}_{K_1K_2}(u)+\mathcal{S}_{K_1K_2}(v),
\end{equation}
and coupled by ordinary $\mathrm{SU}(2)$ Clebsch--Gordan coefficients.

\textit{The amplitude as a linear functional.} In the spinor formalism, the basic ingredient is the 3-point on-shell amplitudes, and higher-point amplitudes can be constructed using the recursive relation or the bootstrap method~\cite{Britto:2004ap,Britto:2005fq}.

Let us start with three-point amplitudes in the principal bundle. In the particle-state description the 3-point amplitude is $\mathcal M_{\{I\}\{J\}}{}^{\{K\}}\bar\psi_{3\{K\}} \equiv\langle p_3\sigma_3|S|p_1\sigma_1,p_2\sigma_2\rangle$, with three velocities in the manifold $u=p_1/m_1$, $v=p_2/m_2$ and $U=p_3/m_3$. Using the unitary equivalence Eq.~\eqref{eq:Hilbertisomorphism} between particle states and sections of $E_s$, and multiplying the (distributional) matrix element by wave functions and integrating, this becomes the probability amplitude, with $d\Phi_3\equiv\prod_{i=1}^{3}d^3\mathbf p_i/(2E_i)$ and $\delta^{(4)}_{123}\equiv\delta^{(4)}(p_1+p_2-p_3)$,
\begin{equation}\label{eq:amplitude}
\int d\Phi_3\,\psi_1^{\{I\}}\psi_2^{\{J\}}\mathcal M_{\{I\}\{J\}}{}^{\{K\}}\bar\psi_{3\{K\}}\,\delta^{(4)}_{123},
\end{equation}
which is linear in $\psi_1$ and $\psi_2$, antilinear in $\psi_3$.

The amplitude $\mathcal{M}$ is built in two steps. First, the $T_1=T_i(p_3,p_1)$ and $T_2=T_2(p_3,p_2)$ transport the constituents little group indices of the spin parts in $p_1$ and $p_2$, to the ones in $P$, respectively,
\begin{equation}\label{eq:spinpart}
\mathcal M_{\{I_1\}\{I_2\}}{}^{\{I_3\}}
=\rho_{s_1}(T_1)_{\{I_1\}}^{\{A\}}\rho_{s_2}(T_2)_{\{I_2\}}^{\{B\}}\,
\hat{\mathcal M}_{\{A\}\{B\}}{}^{\{I_3\}},
\end{equation}
where $\hat{\mathcal M}$ is an $\mathfrak{su}(2)$ tensor at $p_3/m_3$, and it maps the spins $s_1$ and $s_2$ carried by $\rho_{s_1}(T_1)$ and $\rho_{s_2}(T_2)$ onto the spin $s_3$ through the ordinary Clebsch--Gordan series
\begin{equation}\label{eq:decomposition}
V_{s_1}\otimes V_{s_2}=\bigoplus_{S=|s_1-s_2|}^{s_1+s_2}V_S,\qquad
V_S\otimes V_L\longrightarrow V_{s_3}.
\end{equation}
The space $V_L$ corresponds to the $\mathfrak{su}(2)$ irreducible component of $\hat{\mathcal M}$. Second, every index carried by $\hat{\mathcal M}$ belongs to indices at $p_3$, corresponding to the amplitude $0+0\to L$, whose unique form is determined to be
\begin{equation}\label{eq:orbital}
Y^L\equiv[\mathbf 3|1|\mathbf 3\rangle^{L}
=[P^{(I_1}|p_1|P^{I_2}\rangle\cdots[P^{I_{2L-1}}|p_1|P^{I_{2L})}\rangle,
\end{equation}
with $2L$ totally symmetric indices; in the COM frame it is the spherical harmonic $Y_{LL_z}(\hat{\mathbf n})$. That it carries orbital angular momentum $L$ follows from
\begin{equation}\label{eq:Loperator}
\begin{split}
\mathcal S_{K_1K_2} &=  T_{K_1}{}^{I}T_{K_2}{}^{J}\bigl( \lambda_{\alpha(J}\frac{\partial}{\partial\lambda^{I)}_\alpha} + \text{c.c.}\bigr), \\
\mathcal L_{K_1K_2}&=\mathcal J_{K_1K_2}-\mathcal S_{K_1K_2},\\
\mathcal J_{K_1K_2}&=\lambda_{3(K_1}^{\ \ \alpha}
\frac{\partial}{\partial\lambda_3^{K_2)\alpha}}+\text{c.c.}.
\end{split}
\end{equation}

\medskip
{\textit{  Massive spin-spinor amplitudes.}} Coupling the orbital part to the spin part as in Eq.~\eqref{eq:decomposition}, the master three-point covariant orbital-spin (C$LS$) amplitude is given by
\begin{eqnarray}\label{eq:final3ptampLS}
    &\mathcal{M}^{LS}(1^{\{I'\}},2^{\{J'\}},3_{\{K\}})\propto\qquad\nonumber\\
    &\rho_{s_1\{I\}}^{\{I'\}}\rho_{s_2\{J\}}^{\{J'\}}\,
    C^{s_1\{I\};s_2\{J\}}_{S\{N\}}\,C^{L\{B\};S\{N\}}_{s_3\{K\}}\,
    [\mathbf{3}|1|\mathbf{3}\rangle^L_{\{B\}}\,.
\end{eqnarray}
This expression organizes three-point amplitudes by their orbital-spin content, complementing the spinor structures of Arkani-Hamed, Huang, and Huang (AHH)~\cite{Arkani-Hamed:2017jhn}. In the COM frame, these C$LS$ amplitudes reproduce the orbital-spin couplings used in the Jacob-Wick formalism~\cite{Jacob:1959at,Chung:1993da,Chung:1997jn}. This C$LS$ form is completely covariant under the covariant spin and orbital operator action in Eq.~\eqref{eq:Loperator}, without any boost dependence shown in literature~\cite{Zou:2002ar,Zou:2002yy,Jing:2023rwz,Jing:2024mag}.

In the AHH formalism, the little group space indices are converted into the Lorentz space ones via the spin-spinors of external particles. While in this work, built in little-group space rather than from Lorentz invariants, the C$LS$ amplitudes carry well-defined spin and orbital quantum numbers and form {\it an independent, complete basis}. For example, for basic three massive-gauge-boson amplitudes, one could write down totally eight amplitudes $[\mathbf{12}][\mathbf{23}]\langle\mathbf{31}\rangle$,$[\mathbf{12}]\langle\mathbf{23}\rangle[\mathbf{31}]$,$\langle\mathbf{12}\rangle[\mathbf{23}][\mathbf{31}]$, etc. However, C$LS$ amplitudes only contain seven independent elements with $(L,S)=(1,0)$, $(0,1)$, $(1,1)$, $(2,1)$, $(1,2)$, $(2,2)$, $(3,2)$, which are the combination of the above eight amplitudes. Similarly for all other basic three-point amplitudes.

Fermion--antifermion bilinears illustrate the spin--orbital classification. For an equal-mass pair, such as $e^+ e^-$, in a charge-conjugation eigenstate, $P=(-1)^{L+1}$ and $C=(-1)^{L+S}$. The spatial vector current has $J^{PC}=1^{--}$ and admits $(L,S)=(0,1)$ and $(2,1)$, corresponding to the ${}^3S_1$ and ${}^3D_1$ channels. The pseudoscalar channel has $(L,S)=(0,0)$ and $J^{PC}=0^{-+}$. Near threshold, these include the leading ortho- and parapositronium configurations, respectively. In literature~\cite{Jacob:1959at,Chung:1993da,Chung:1997jn}, the $LS$ structure is only hold in the COM frame, however, in this work both $J^{PC}$ and $LS$ hold for any momenta. The supplementary material gives the bilinear decomposition. This can be extended to covariant multipole expansion for general form factors~\cite{Huang:2026kfq}.

Given 3-point C$LS$ amplitude basis, higher-point amplitudes can be constructively obtained by recursive relation and bootstrap method~\cite{Britto:2004ap,Britto:2005fq}. The advantage of this C$LS$ amplitudes is that it is the partial wave eigen-states with covariant orbital and spin eigen-values. Therefore, partial wave analysis can be performed covariantly, without the need of boosting back to the successive COM frames in the decay-chains. This covariant partial-wave analysis has been applied~\cite{Huang:2026egv} to $\Lambda_c\to\Lambda\pi^+\pi^0$, to help identify all the immediate particles: $\rho(770)^{+}$, $\Sigma(1385)^{+}$,
$\Sigma(1385)^{0}$, $\Sigma(1670)^{+}$, $\Sigma(1670)^{0}$,
$\Sigma(1750)^{+}$, and $\Sigma(1750)^{0}$~\cite{BESIII:2022udq, tfpwa}.

The above C$LS$ amplitudes can be extended~\footnote{In principle, massless particles have a different momentum orbit and little group, so their bundle construction must be treated separately. However, in the spinor formalism, we find the massive partial wave decomposition can be extended to amplitudes with massless particles.} to the ones with massless constituents. When particle 2 becomes massless, both $S(p_2)$ and $T_2$ are absent; instead the helicity of particle 2 couples to orbital momentum, $\mathcal{Y}^L=[2\mathbf{P}]^{2|h_2|}[\mathbf{P}|1|\mathbf{P}\rangle^L$ or $\mathcal{Y}^L=\langle2\mathbf{P}\rangle^{2|h_2|}[\mathbf{P}|1|\mathbf{P}\rangle^L$ depending on the sign of $h_2$. The full amplitude couples this orbital factor to the spin sector of particle 1, described by $\tau_1^{2s_1}$: $M=C^{\{J\}\{I'\}}_{\{K\}}\mathcal{Y}^L_{\{J\}}\tau_{1\{I'\}}^{2s\{I\}}$. If both constituents become massless, the amplitude is dictated by the spin of the massive final state: $M=[1\mathbf{P}]^{J+h_1-h_2}[2\mathbf{P}]^{J-h_1+h_2}/[12]^{J-h_1-h_2}$.

\textbf{\textit{Summary and Outlook.}} By identifying the spin fiber and parallel transport over the velocity manifold, we have constructed a new covariant spin operator, which couples different particle's spins in a Lorentz covariant way. The resulting three-point C$LS$ structures organize amplitudes by orbital and spin angular momentum while retaining an explicit prescription for comparing spin frames. Therefore, by introducing the covariant spin and orbital operators in relativistic quantum theory, there is no mixture between the covariant spin and orbital structures under the Lorentz boost. As application, higher-point C$LS$ amplitude, as the partial wave basis, can be constructed from the three-point C$LS$ amplitude basis, which can be applied to high energy scattering, hadronic physics, nucleon structure, and gravitational wave calculation, etc.

\bibliography{ref}

@article{Eguchi:1980jx,
    author = "Eguchi, Tohru and Gilkey, Peter B. and Hanson, Andrew J.",
    title = "{Gravitation, Gauge Theories and Differential Geometry}",
    reportNumber = "PRINT-80-0182 (LBL,BERKELEY)",
    doi = "10.1016/0370-1573(80)90130-1",
    journal = "Phys. Rept.",
    volume = "66",
    pages = "213",
    year = "1980"
}

@inproceedings{Dixon:2013uaa,
    author = "Dixon, Lance J.",
    title = "{A brief introduction to modern amplitude methods}",
    booktitle = "{TASI in Elementary Particle Physics}",
    eprint = "1310.5353",
    archivePrefix = "arXiv",
    primaryClass = "hep-ph",
    reportNumber = "SLAC-PUB-15775",
    doi = "10.5170/CERN-2014-008.31",
    pages = "31--67",
    year = "2014"
}

@article{Witten:2003nn,
    author = "Witten, Edward",
    title = "{Perturbative gauge theory as a string theory in twistor space}",
    eprint = "hep-th/0312171",
    archivePrefix = "arXiv",
    doi = "10.1007/s00220-004-1187-3",
    journal = "Commun. Math. Phys.",
    volume = "252",
    pages = "189--258",
    year = "2004"
}

@article{Benincasa:2007xk,
    author = "Benincasa, Paolo and Cachazo, Freddy",
    title = "{Consistency Conditions on the S-Matrix of Massless Particles}",
    eprint = "0705.4305",
    archivePrefix = "arXiv",
    primaryClass = "hep-th",
    reportNumber = "UWO-TH-07-09",
    month = "5",
    year = "2007"
}

@article{Britto:2005fq,
    author = "Britto, Ruth and Cachazo, Freddy and Feng, Bo and Witten, Edward",
    title = "{Direct proof of tree-level recursion relation in Yang-Mills theory}",
    eprint = "hep-th/0501052",
    archivePrefix = "arXiv",
    doi = "10.1103/PhysRevLett.94.181602",
    journal = "Phys. Rev. Lett.",
    volume = "94",
    pages = "181602",
    year = "2005"
}

@article{Britto:2004ap,
    author = "Britto, Ruth and Cachazo, Freddy and Feng, Bo",
    title = "{New recursion relations for tree amplitudes of gluons}",
    eprint = "hep-th/0412308",
    archivePrefix = "arXiv",
    doi = "10.1016/j.nuclphysb.2005.02.030",
    journal = "Nucl. Phys. B",
    volume = "715",
    pages = "499--522",
    year = "2005"
}

@inproceedings{Cheung:2017pzi,
    author = "Cheung, Clifford",
    title = "{TASI lectures on scattering amplitudes.}",
    booktitle = "{TASI in Elementary Particle Physics}",
    eprint = "1708.03872",
    archivePrefix = "arXiv",
    primaryClass = "hep-ph",
    reportNumber = "CALT-TH-2017-041",
    doi = "10.1142/9789813233348_0008",
    pages = "571--623",
    year = "2018"
}

@book{KobayashiNomizu:1969,
    author    = "Kobayashi, Shoshichi and Nomizu, Katsumi",
    title     = "{Foundations of Differential Geometry. Volume II}",
    publisher = "Interscience Publishers",
    address   = "New York",
    year      = "1969",
    series    = "Interscience Tracts in Pure and Applied Mathematics",
    number    = "15",
    isbn      = "978-0-470-49648-0"
}

@book{Dubrovin:1985,
    author    = "Dubrovin, B. A. and Fomenko, A. T. and Novikov, S. P.",
    title     = "{Modern Geometry --- Methods and Applications: Part II}",
    series    = "Graduate Texts in Mathematics",
    volume    = "104",
    publisher = "Springer",
    year      = "1985",
    doi       = "10.1007/978-1-4612-1100-6"
}

@article{Thomas:1926dy,
    author = "Thomas, L. H.",
    title = "{The motion of a spinning electron}",
    doi = "10.1038/117514a0",
    journal = "Nature",
    volume = "117",
    pages = "514",
    year = "1926"
}

@article{Aharonov:1987gg,
    author = "Aharonov, Y. and Anandan, J.",
    title = "{Phase Change During a Cyclic Quantum Evolution}",
    reportNumber = "Print-87-0207 (SOUTH CAROLINA)",
    doi = "10.1103/PhysRevLett.58.1593",
    journal = "Phys. Rev. Lett.",
    volume = "58",
    pages = "1593",
    year = "1987"
}

@article{Berry:1984jv,
    author = "Berry, Michael V.",
    title = "{Quantal phase factors accompanying adiabatic changes}",
    doi = "10.1098/rspa.1984.0023",
    journal = "Proc. Roy. Soc. Lond. A",
    volume = "392",
    pages = "45--57",
    year = "1984"
}

@article{Wigner:1939cj,
    author = "Wigner, Eugene P.",
    editor = "Kim, Y. S. and Zachary, W. W.",
    title = "{On Unitary Representations of the Inhomogeneous Lorentz Group}",
    doi = "10.2307/1968551",
    journal = "Annals Math.",
    volume = "40",
    pages = "149--204",
    year = "1939"
}

@article{Polyzou:2012ut,
    author = {Polyzou, W. N. and Gl{\"o}ckle, W. and Witala, H.},
    title = "{Spin in relativistic quantum theory}",
    eprint = "1208.5840",
    archivePrefix = "arXiv",
    primaryClass = "nucl-th",
    doi = "10.1007/s00601-012-0526-8",
    journal = "Few Body Syst.",
    volume = "54",
    pages = "1667--1704",
    year = "2013"
}

@article{Newton:1949cq,
    author = "Newton, T. D. and Wigner, Eugene P.",
    title = "{Localized States for Elementary Systems}",
    doi = "10.1103/RevModPhys.21.400",
    journal = "Rev. Mod. Phys.",
    volume = "21",
    pages = "400--406",
    year = "1949"
}

@article{Palge:2023mmq,
    author = "Palge, Veiko and Pfeifer, Christian",
    title = "{Thomas-Wigner rotation as a holonomy for spin-1/2 particles}",
    eprint = "2310.08121",
    archivePrefix = "arXiv",
    primaryClass = "quant-ph",
    doi = "10.1103/PhysRevA.109.032206",
    journal = "Phys. Rev. A",
    volume = "109",
    number = "3",
    pages = "032206",
    year = "2024"
}

@article{Oblak:2017oge,
    author = "Oblak, Blagoje",
    title = "{Probing Wigner Rotations for Any Group}",
    eprint = "1710.06883",
    archivePrefix = "arXiv",
    primaryClass = "hep-th",
    doi = "10.1016/j.geomphys.2018.03.008",
    journal = "J. Geom. Phys.",
    volume = "129",
    pages = "168--185",
    year = "2018"
}

@article{Huang:2026kfq,
    author = "Huang, Hong and Tan, Tuo and Wang, Yi-Ning and Yu, Jiang-Hao",
    title = "{Covariant Spinor Formalism for Multipole Expanded Form Factor}",
    eprint = "2605.04030",
    archivePrefix = "arXiv",
    primaryClass = "hep-ph",
    month = "5",
    year = "2026"
}

@article{Huang:2026egv,
    author = "Huang, Hong and Wang, Yi-Ning and Yu, Jiang-Hao",
    title = "{Covariant canonical-spinor amplitudes for partial wave analysis}",
    eprint = "2603.04487",
    archivePrefix = "arXiv",
    primaryClass = "hep-ph",
    month = "3",
    year = "2026"
}

@article{Arkani-Hamed:2017jhn,
    author = "Arkani-Hamed, Nima and Huang, Tzu-Chen and Huang, Yu-tin",
    title = "{Scattering amplitudes for all masses and spins}",
    eprint = "1709.04891",
    archivePrefix = "arXiv",
    primaryClass = "hep-th",
    reportNumber = "NCTS-TH/1714, NCTS-TH-1714",
    doi = "10.1007/JHEP11(2021)070",
    journal = "JHEP",
    volume = "11",
    pages = "070",
    year = "2021"
}

@article{Jing:2024mag,
    author = "Jing, Hao-Jie and Wu, Shu-Ming and Wu, Jia-Jun",
    title = "{Algorithms for partial wave amplitudes in the covariant L-S scheme}",
    eprint = "2405.06576",
    archivePrefix = "arXiv",
    primaryClass = "hep-ph",
    doi = "10.1103/PhysRevD.110.016014",
    journal = "Phys. Rev. D",
    volume = "110",
    number = "1",
    pages = "016014",
    year = "2024"
}

@article{Jacob:1959at,
    author = "Jacob, M. and Wick, G. C.",
    title = "{On the General Theory of Collisions for Particles with Spin}",
    doi = "10.1006/aphy.2000.6022",
    journal = "Annals Phys.",
    volume = "7",
    pages = "404--428",
    year = "1959"
}

@book{Weinberg:1995mt,
    author = "Weinberg, Steven",
    title = "{The Quantum theory of fields. Vol. 1: Foundations}",
    doi = "10.1017/CBO9781139644167",
    isbn = "978-0-521-67053-1, 978-0-511-25204-4",
    publisher = "Cambridge University Press",
    month = "6",
    year = "2005"
}

@article{Jing:2023rwz,
    author = "Jing, Hao-Jie and Ben, Di and Wu, Shu-Ming and Wu, Jia-Jun and Zou, Bing-Song",
    title = "{Covariant orbital-spin scheme for any spin based on irreducible tensor}",
    eprint = "2301.01575",
    archivePrefix = "arXiv",
    primaryClass = "hep-ph",
    doi = "10.1007/JHEP06(2023)039",
    journal = "JHEP",
    volume = "06",
    pages = "039",
    year = "2023"
}

@article{Zou:2002yy,
    author = "Zou, B. S. and Hussain, F.",
    title = "{Covariant L-S scheme for the effective N*NM couplings}",
    eprint = "hep-ph/0210164",
    archivePrefix = "arXiv",
    reportNumber = "BIHEP-TH-2002-29",
    doi = "10.1103/PhysRevC.67.015204",
    journal = "Phys. Rev. C",
    volume = "67",
    pages = "015204",
    year = "2003"
}

@article{Zou:2002ar,
    author = "Zou, B. S. and Bugg, D. V.",
    title = "{Covariant tensor formalism for partial wave analyses of psi decay to mesons}",
    eprint = "hep-ph/0211457",
    archivePrefix = "arXiv",
    doi = "10.1140/epja/i2002-10135-4",
    journal = "Eur. Phys. J. A",
    volume = "16",
    pages = "537--547",
    year = "2003"
}

@article{Chung:1997jn,
    author = "Chung, S. U.",
    title = "{A General formulation of covariant helicity coupling amplitudes}",
    reportNumber = "BNL-QGS-97-051",
    doi = "10.1103/PhysRevD.57.431",
    journal = "Phys. Rev. D",
    volume = "57",
    pages = "431--442",
    year = "1998"
}

@article{Chung:1993da,
    author = "Chung, S. U.",
    title = "{Helicity coupling amplitudes in tensor formalism}",
    doi = "10.1103/PhysRevD.56.4419",
    journal = "Phys. Rev. D",
    volume = "48",
    pages = "1225--1239",
    year = "1993",
    note = "[Erratum: Phys.Rev.D 56, 4419 (1997)]"
}

@article{Dittmaier:1998nn,
    author = "Dittmaier, Stefan",
    title = "{Weyl-van der Waerden formalism for helicity amplitudes of massive particles}",
    eprint = "hep-ph/9805445",
    archivePrefix = "arXiv",
    reportNumber = "CERN-TH-98-143",
    doi = "10.1103/PhysRevD.59.016007",
    journal = "Phys. Rev. D",
    volume = "59",
    pages = "016007",
    year = "1998"
}

@article{Habermann:2024sxs,
    author = "Habermann, Kai and Mikhasenko, Mikhail",
    title = "{Wigner rotations for cascade reactions}",
    eprint = "2409.06913",
    archivePrefix = "arXiv",
    primaryClass = "hep-ph",
    doi = "10.1103/PhysRevD.111.056015",
    journal = "Phys. Rev. D",
    volume = "111",
    number = "5",
    pages = "056015",
    year = "2025"
}

@article{Callan:1969sn,
    author = "Callan, Jr., Curtis G. and Coleman, Sidney R. and Wess, J. and Zumino, Bruno",
    title = "{Structure of phenomenological Lagrangians. 2.}",
    doi = "10.1103/PhysRev.177.2247",
    journal = "Phys. Rev.",
    volume = "177",
    pages = "2247--2250",
    year = "1969"
}

@article{Coleman:1969sm,
    author = "Coleman, Sidney R. and Wess, J. and Zumino, Bruno",
    title = "{Structure of phenomenological Lagrangians. 1.}",
    doi = "10.1103/PhysRev.177.2239",
    journal = "Phys. Rev.",
    volume = "177",
    pages = "2239--2247",
    year = "1969"
}

@article{Badger:2023eqz,
    author = "Badger, Simon and Henn, Johannes and Plefka, Jan Christoph and Zoia, Simone",
    title = "{Scattering Amplitudes in Quantum Field Theory}",
    eprint = "2306.05976",
    archivePrefix = "arXiv",
    primaryClass = "hep-th",
    doi = "10.1007/978-3-031-46987-9",
    journal = "Lect. Notes Phys.",
    volume = "1021",
    pages = "pp.",
    year = "2024"
}

@article{Elvang:2015rqa,
    author = "Elvang, Henriette and Huang, Yu-tin",
    title = "{Scattering Amplitudes in Gauge Theory and Gravity}",
    publisher = "Cambridge University Press",
    year = "2015",
    isbn = "978-1-316-19142-2"
}

@article{BESIII:2022udq,
    author = "Ablikim, Medina and others",
    collaboration = "BESIII",
    title = "{Partial wave analysis of the charmed baryon hadronic decay $ {\Lambda}_c^{+} ${\textrightarrow} {\ensuremath{\Lambda}}{\ensuremath{\pi}}$^{+}${\ensuremath{\pi}}$^{0}$}",
    eprint = "2209.08464",
    archivePrefix = "arXiv",
    primaryClass = "hep-ex",
    doi = "10.1007/JHEP12(2022)033",
    journal = "JHEP",
    volume = "12",
    pages = "033",
    year = "2022"
}

@article{tfpwa,
title = {Open-source framework {TF-PWA} package},
journal = {GitHub link: https://github.com/jiangyi15/tf-pwa},
year = {2020},
url = {https://github.com/jiangyi15/tf-pwa},
author = {Y. Jiang and others}
}

\newpage

\appendix

\section{Supplementary Materials}

\medskip
\textbf{\textit{The associated bundle, its sections and frames.}} Let $V_s\cong\mathbb C^{2s+1}$ carry the irreducible unitary representation $\rho_s$ of the little group $H=\mathrm{SU}(2)$, and let $G$ be the principal $H$-bundle over the velocity manifold $M$, with projection $\pi(g)=gu_\star$. For $g\in G$ and $\psi\in V_s$ the associated bundle is the set of pairs $(g,\psi)$ modulo the identification
\begin{equation}\label{eq:sm-associatedbundle}
\begin{split}
E_s\equiv G\times_H V_s=(G\times V_s)/\!\sim,\\
(gh,\psi)\sim(g,\rho_s(h)\psi)\qquad\forall h\in H,
\end{split}
\end{equation}
where an element of $E_s$ is written $(g,\psi)$, the pair and its class being identified in the notation; it lies over $u=\pi(g)$ and carries the spin state $\psi\in V_s$. The spin-$s$ field is a section of this bundle, that is, a smooth map $\Psi:M\to E_s$ with $\pi_{E_s}\circ\Psi=\mathrm{id}_M$. Once a section $L_u$ of $G$ is chosen, any element over $u$ reads $g=L_uh$ with $h\in H$, so every class over $u$ has the representative $(L_u,\rho_s(h)\psi)$ and the fiber $(E_s)_u\cong V_s$ is the spin space at velocity $u$; a section is then the same thing as a map $\psi:M\to V_s$, through $\Psi(u)=(L_u,\psi(u))$, and replacing $L_u$ by the equally admissible lift $L_u'=L_uh(u)$, $h:M\to H$, changes it according to
\begin{equation}\label{eq:sm-gauge}
\psi(u)\ \longrightarrow\ \rho_s\big(h(u)\big)^{-1}\psi(u),
\end{equation}
since $[L_uh(u),\psi'(u)]=[L_u,\rho_s(h(u))\psi'(u)]$ must equal $(L_u,\psi(u))$; this is Eq.~\eqref{eq:sm-gauge}, the little-group redefinition of the spin axis.

Fixing an orthonormal basis $\{e_\sigma\}_{\sigma=-s}^{s}$ of $V_s$ (the representation $\rho_s$ being taken in the standard basis with respect to it), the same trivialization provides a global frame of $E_s$,
\begin{equation}\label{eq:sm-frame}
b_\sigma(u)\equiv[L_u,e_\sigma],\qquad \Psi(u)=\psi^\sigma(u)\,b_\sigma(u),
\end{equation}
with $b_\sigma(u)$ corresponding to the Wigner state $|mu,\sigma\rangle=U(L_u)|k,\sigma\rangle$, $k=mu_\star$, which is the distributional section localized at $u$; the choice of $L_u$ is precisely the choice of its rest-frame axes.

\medskip
The one-particle Hilbert space is the space of square-integrable sections,
\begin{equation}\label{eq:sm-hilbert}
\begin{split}
\mathcal H_s=\Big\{\Psi\in\Gamma(E_s)\ \Big|\ \langle\Psi,\Psi\rangle<\infty\Big\},\\
\langle\Psi_1,\Psi_2\rangle=\int_M\frac{d^3\mathbf u}{u^0}\,\big(\psi_1(u),\psi_2(u)\big)_{V_s},
\end{split}
\end{equation}
where $d^3\mathbf u/u^0$ is the Lorentz-invariant measure on $M$ and $(\,\ )_{V_s}$ the $H$-invariant Hermitian form on $V_s$. The integrand is independent of the chosen $L_u$: by Eq.~\eqref{eq:sm-gauge} both representatives acquire $\rho_s(h)^{-1}$, and
\begin{equation}\label{eq:sm-invariance}
\big(\rho_s(h)^{-1}\psi_1,\rho_s(h)^{-1}\psi_2\big)_{V_s}=\big(\psi_1,\psi_2\big)_{V_s}
\end{equation}
by unitarity of $\rho_s$. In a fixed trivialization, Eq.~\eqref{eq:sm-hilbert} gives the unitary isomorphism of Eq.~\eqref{eq:Hilbertisomorphism}. The particle-state description is the more basic of the two: the states $|mu,\sigma\rangle$ are distributional, $\delta$-function normalized, and need not be normalizable, whereas $\mathcal H_s$ contains only normalizable states; the two are nevertheless physically equivalent, agreeing on all matrix elements of bounded observables, and we pass freely between them.

The left action of $G$ on itself descends to the associated bundle,
\begin{equation}\label{eq:sm-leftaction}
\Lambda\cdot(L_u,\psi)\equiv(\Lambda L_u,\psi)=(L_{\Lambda u},\rho_s(W)\psi),
\end{equation}
thus Lorentz transformation $\Lambda$ carries $\psi$ into a new wave function $\psi'$, whose value at the velocity $\Lambda u$ is $\psi'(\Lambda u)=\rho_s\big(W(\Lambda,u)\big)\psi(u)$; its value at $u$ is therefore obtained with $u\to\Lambda^{-1}u$,
\begin{equation}\label{eq:sm-inducedrep}
\big(U(\Lambda)\psi\big)(u)=\rho_s\big(W(\Lambda,\Lambda^{-1}u)\big)\psi(\Lambda^{-1}u),
\end{equation}
 On the particle states the same operator reads $U(\Lambda)|mu,\sigma\rangle=\sum_{\sigma'}|m\Lambda u,\sigma'\rangle\rho_s(W(\Lambda,u))^{\sigma'}{}_{\sigma}$, which is the standard Wigner formula; the two forms are identified by the isomorphism of Eq.~\eqref{eq:Hilbertisomorphism}. 

\medskip
\textbf{\textit{Spin as the generator of the right action.}} For $S_{IJ}\in\mathfrak{su}(2)$ the derived representation on the fiber is
\begin{equation}\label{eq:sm-representation}
\rho_s(S_{IJ})\equiv\frac{d}{dt}\Big|_{0}\rho_s\big(e^{tS_{IJ}}\big)\in\mathrm{End}(V_s).
\end{equation}
On the principal bundle the same element generates the vertical vector fields of Eq.~\eqref{eq:Sprincipal} of the main text, $S_I{}^J=\lambda^{(J}_\alpha\partial/\partial\lambda^{I)}_\alpha+\text{c.c.}$ and $S_\rho=\tfrac12(\sigma_\rho)^I{}_J S^J{}_I$, which close on $\mathfrak{su}(2)$ and move $g$ along its fiber, since $e^{tS_{IJ}}\in H$ stabilizes $u_\star$. Carried to every fiber by the trivialization of Eq.~\eqref{eq:sm-associatedbundle}, the same element acts through $\rho_s'$, yielding the fiber-wise generator
\begin{equation}\label{eq:sm-Sinduced}
\mathcal S_{IJ}\in\mathrm{End}(E_s),\qquad \mathcal S_{IJ}(L_u,\psi)=(L_u,\rho_s(S_{IJ})\psi),
\end{equation}
so that in the same trivialization $\mathcal S_{IJ}\psi(u)=\rho_s(S_{IJ})\psi(u)$ for every $S_{IJ}\in\mathfrak{su}(2)$. Its components along the spherical basis $S_\rho$, normalized by $\rho_{1/2}(S_\rho)=\tfrac12\sigma_\rho$, are the fiber-wise operators of the main text. Since the trivialization is built from $L_u$, the operator depends on the choice of standard boosts and is not canonically defined on $E_s$.

The right action on $G$ does not descend to $E_s$: the prescription $(g,\psi)\mapsto(gh,\psi)$ is not well defined, because applied to the equivalent representative $(g\tilde h,\rho_s(\tilde h)^{-1}\psi)$ it gives
\begin{equation}\label{eq:sm-notdescend}
(g\tilde hh,\rho_s(\tilde h)^{-1}\psi)=(g,\rho_s(\tilde h)\rho_s(h)\rho_s(\tilde h)^{-1}\psi),
\end{equation}
which differs from $(gh,\psi)=(g,\rho_s(h)\psi)$ unless $\rho_s$ is conjugation invariant, i.e. unless $H$ is Abelian. What the right action induces instead is an action on the trivialization itself, $L_u\to L_uh(u)$, whose infinitesimal form is Eq.~\eqref{eq:sm-Sinduced}. Consequently $\mathcal S_{IJ}$ is gauge covariant rather than gauge invariant: under $L_u\to L_uh(u)$ one has $S_\rho\mapsto \rho_1(h)_\rho{}^\sigma S_\sigma$. The corresponding Hilbert-space operator acts pointwise:
\begin{equation}\label{eq:sm-pointwise}
(S_\rho\Psi)(u)=S_\rho(u)\,\Psi(u).
\end{equation}

As a cross-check that $S_\rho$ is indeed the spin operator, note that it has a covariant realization in terms of the Pauli--Lubanski vector. The polarization vectors are those of the section $L_u$,
\begin{equation}\label{eq:polarization}
\epsilon^\mu_\rho(u)\equiv[L_u]^\mu{}_\nu\,(0,\mathbf e_\rho)^\nu,\qquad \rho=\pm1,0,
\end{equation}
built on the spherical basis $\mathbf e_{\pm1}=\tfrac{1}{\sqrt2}(\mathbf e_x\pm i\mathbf e_y)$, $\mathbf e_0=\mathbf e_z$. At the rest velocity $u_\star$ one has $L_{u_\star}=\mathbf 1$, so Eq.~\eqref{eq:polarization} gives $\epsilon^\mu_\rho(u_\star)=(0,\mathbf e_\rho)$, while the Pauli--Lubanski vector there is $\mathcal W_\mu=(0,m\mathbf J)$ by Eq.~\eqref{eq:Spoincare}. The covariant operator
Writing the covariant form of the operator as $S_\rho(u)\equiv\frac{1}{m}\epsilon^\mu_\rho(u)\mathcal W_\mu(u)$, one therefore has
\begin{equation}\label{eq:SpinPauliLubanski}
S_\rho(u_\star)=\mathbf e_\rho\cdot\mathbf J=\mathcal J_\rho ,
\end{equation}
the spherical component of the spin; this $S_\rho$ is thus indeed the spin operator.

\medskip

\textbf{\textit{The transport operator, Wigner rotation and holonomy.}} In above the spin is defined at a single velocity. To compare two such spins one must carry the frame from one velocity to another via transport.

\textit{The horizontal lift and the transport.} The connection $\omega_h$ specifies which curves in $G$ are horizontal, lifting a path $\gamma$ in $M$ to a path $\tilde\gamma$ in $G$, and the frame that is carried along this lift is what we call the transport $T$. Its horizontal lift is the unique curve $\tilde\gamma$ with $\pi\circ\tilde\gamma=\gamma$ whose tangent is horizontal, $\omega_h(\dot{\tilde\gamma})=0$. Starting at the frame $L_u$, it ends at a lift of $v$, that is, at $L_v$ times an element of $H$; that element is $T(v,u)$.
\begin{equation}\label{eq:sm-lift}
\tilde\gamma(1)=L_v\,T_\gamma,\qquad T_\gamma\in H,
\end{equation}
and $T_\gamma=\mathcal P\exp(-\int_\gamma\omega_h)$ of Eq.~\eqref{eq:generictransport} is the integrated form of the horizontality condition. The horizontal directions are the left translates of $\mathfrak m$; the lift is unique once its starting point is fixed, and a different starting point changes $T$ by the right multiplication that relates the two frames.

\textit{The Wilson line and the Wilson loop.} For a general path the transport of Eq.~\eqref{eq:generictransport} is the Wilson line of the connection,
\begin{equation}\label{eq:sm-wilsonline}
W[\gamma(\Lambda,u)]=\mathcal P\exp\Big(-\int_{\gamma(\Lambda,u)}\omega_h\Big),
\end{equation}
formally identical to the Wilson line of a gauge field, the connection $\omega_h$ playing the role of the gauge potential. Applied to the motion of an accelerated particle it reproduces the Thomas precession~\cite{Thomas:1926dy,Oblak:2017oge}. When the open path $\gamma(\Lambda,u)$ is closed into a loop~\cite{Palge:2023mmq} by adjoining the canonical boost paths $\gamma_u$ and $\gamma_{\Lambda u}^{-1}$, which the canonical gauge trivializes, the Wilson line becomes the holonomy of a closed loop, i.e.\ a Wilson loop, and the Wigner rotation is the geometric phase~\cite{Berry:1984jv,Aharonov:1987gg} of the triangle $u_\star\to u\to\Lambda u\to u_\star$. The construction holds for any section. The transport is therefore a Wigner rotation, the rotation that $\Lambda$ induces at $u$ and that Eq.~\eqref{eq:Spintransform} of the main text carries: the trivialised form of the transport, not an independent ingredient.
An arbitrary path gives a well-defined transport, but it depends on the whole path rather than on its two endpoints.

\medskip
\textit{Why the path is a geodesic.} Not every curve is admissible: a path cannot connect two different mass shells, so it is defined on the common velocity manifold $M$. The covariance condition Eq.~\eqref{eq:pathequivariance} is what singles out the geodesic. Take a Lorentz transformation $\Lambda_0$ that leaves both $u$ and $v$ invariant; it lies in their common stabiliser, and Eq.~\eqref{eq:pathequivariance} then reduces to $\gamma(v,u)=\Lambda_0\,\gamma(v,u)$, so every point of the curve is invariant under that stabiliser. In the rest frame of $u$ the stabiliser of the pair is the $\mathrm{SO}(2)$ of rotations about the direction of $\mathbf v$, and a curve from $u$ to $v$ left pointwise invariant by it must be the straight line joining them.

Parameterise that line by $s\in[0,1]$. In the rest frame of $u$, where $u=(1,\mathbf 0)$ and $v=(\gamma,\mathbf v)$, its points are
\begin{equation*}
\frac{\big((1-s)+s\gamma,\ s\mathbf v\big)}{\sqrt{(1-s)^2+s^2+2s(1-s)\gamma}},
\end{equation*}
the denominator normalising the four-velocity; in a general frame this reads $\gamma(s)=q(s)/\sqrt{q^2(s)}$ with $q(s)=(1-s)u+sv$, which is Eq.~\eqref{eq:geodesic}. Since $(1-s)\Lambda u+s\Lambda v=\Lambda q(s)$, the covariance condition Eq.~\eqref{eq:pathequivariance} holds automatically, so this is the geodesic of $M$.

\textit{The transport along a geodesic and the boost.} With the connection $\omega_h$, the horizontal lift of the geodesic Eq.~\eqref{eq:geodesic} through the frame $L_u$ is $\tilde\gamma(t)=L_u\,e^{t\xi}$, indeed $\tilde\gamma^{-1}\dot{\tilde\gamma}=\xi\in\mathfrak m$, so that
\begin{equation}\label{eq:sm-geodesictransport}
T(v,u)=L_v^{-1}L_u\,e^{\xi},
\end{equation}
with $B(v,u)\equiv L_u\,e^{\xi}L_u^{-1}$ the boost along the geodesic. Taking $\zeta$ to be a pure boost, $e^{\zeta}=B_c(v')$ with $v'\equiv L_u^{-1}v$, gives $B(v,u)=L_u\,B_c(v')\,L_u^{-1}$; $B(v,u)$ therefore satisfies Lorentz covariance, $\Lambda B(v,u)\Lambda^{-1}=B(\Lambda v,\Lambda u)$, and is fixed as
\begin{equation}\label{eq:sm-geodesicboost}
B(v,u)=\frac{\delta_\alpha{}^\beta+v_{\alpha\dot\alpha}u^{\dot\alpha\beta}}{\sqrt{2(1+u\cdot v)}}.
\end{equation}
Substituting it into Eq.~\eqref{eq:sm-geodesictransport} gives the transport in closed form,
\begin{equation}\label{eq:sm-Tclosed}
T(v,u)=L_v^{-1}\,B(v,u)\;L_u ,
\end{equation}
which is the spinor expression of Eq.~\eqref{eq:Ttwostructures} of the main text, written there with one index at each leg. For this boost the Wigner rotation is defined by the standard relation $B(v,u)\,L_u=L_v\,W(B(v,u),u)$, that is $W(\Lambda,u)\equiv L_{\Lambda u}^{-1}\Lambda L_u$ with $\Lambda u=v$; since $B(v,u)L_u=L_ue^{\xi}=L_vT(v,u)$, the two coincide, $T(v,u)=W(B(v,u),u)$, so the transport is the Wigner rotation of the boost and not an independent ingredient.

The properties usually imposed on $T$ now follow rather than being assumed: $T^\dagger T=\mathbf 1$, $\det T=1$ and $T(u,u)=\mathbf 1$ because $T\in H$, and also from the spinor form Eq.~\eqref{eq:Ttwostructures} of the main text; $T(v,u)=T(u,v)^{-1}$ from Eq.~\eqref{eq:sm-Tclosed}; the gauge law $T(v,u)\mapsto h(v)^{-1}T(v,u)h(u)$ under $L_u\to L_uh(u)$; and $T(\Lambda v,\Lambda u)=W(\Lambda,v)\,T(v,u)\,W(\Lambda,u)^{-1}$, which is Eq.~\eqref{eq:symOfT} of the main text, from the $G$-invariance of the connection, the same invariance that makes the transport depend on the endpoints only.

\medskip
\textbf{\textit{Three-point C$LS$ amplitudes.}} We consider the processes $s_3\to s_1+s_2$ with each $s_i$ at most one, writing $\phi$, $f$ and $V$ for a scalar, a fermion and a vector boson. Their amplitudes are constructed in Ref.~\cite{Arkani-Hamed:2017jhn} and collected in Table~\ref{tab:3ptstructures}. That construction does not distinguish incoming from outgoing particles, whereas for our C$LS$ amplitudes exchanging the incoming and outgoing pairs gives different amplitudes: for $ffV$, for instance, the C$LS$ amplitudes can be organised either as $V\to ff$ or as $f\to fV$, because the $LS$ quantum numbers are defined with respect to a chosen pair. We take $LS$ to be the orbital and spin quantum numbers of the outgoing pair.

\begin{table}[t]
\centering
\footnotesize
\begin{tabular}{@{}ll@{}}
\toprule
external particles & 3-point amplitude structure \\
\midrule
$(1^{1/2},2^{1/2},3^{1})$
& $[\mathbf{23}]\langle\mathbf{31}\rangle$,\quad$\langle\mathbf{23}\rangle[\mathbf{31}]$,\quad
  $\langle\mathbf{23}\rangle\langle\mathbf{31}\rangle$,\quad
  $[\mathbf{23}][\mathbf{31}]$ \\
$(1^{1/2},2^{1/2},3^{0})$
& $[\mathbf{12}]$,\quad $\langle\mathbf{12}\rangle$ \\
$(1^{1},2^{1},3^{1})$
& $[\mathbf{12}][\mathbf{23}]\langle\mathbf{31}\rangle$,\quad
  $[\mathbf{12}]\langle\mathbf{23}\rangle[\mathbf{31}]$,\quad
  $\langle\mathbf{12}\rangle[\mathbf{23}][\mathbf{31}]$,\\
& $[\mathbf{12}]\langle\mathbf{23}\rangle\langle\mathbf{31}\rangle$,\quad
  $\langle\mathbf{12}\rangle\langle\mathbf{23}\rangle[\mathbf{31}]$,\quad
  $[\mathbf{12}][\mathbf{23}][\mathbf{31}]$,\\
& $\langle\mathbf{12}\rangle\langle\mathbf{23}\rangle\langle\mathbf{31}\rangle$ \\
$(1^{1},2^{1},3^{0})$
& $\langle\mathbf{12}\rangle[\mathbf{12}]$,\quad
  $\langle\mathbf{12}\rangle^{2}$,\quad
  $[\mathbf{12}]^{2}$ \\
$(1^{1},2^{0},3^{0})$
& $\langle\mathbf{3}|p_1|\mathbf{3}]$ \\
$(1^{0},2^{0},3^{0})$
& $1$ \\
\bottomrule
\end{tabular}
\caption{Independent three-point amplitude structures for the indicated external particles.}
\label{tab:3ptstructures}
\end{table}

With Eq.~\eqref{eq:final3ptampLS} the three-point C$LS$ amplitudes are constructed directly, as follows. For three scalars, $s_1=s_2=s_3=0$, there is the single combination $(L,S)=(0,0)$,
\begin{equation}\label{eq:sm-SSS}
\mathcal M(0,0)\propto 1 .
\end{equation}

\noindent For $f\to f\phi$, that is $s_3=\tfrac12$ with $s_1=\tfrac12$ and $s_2=0$, the two combinations $(0,\tfrac12)$ and $(1,\tfrac12)$ give
\begin{equation}\label{eq:sm-FfS}
\begin{array}{@{}l@{}}
\mathcal M(0,\tfrac12)\propto\big(\langle\mathbf{13}\rangle-[\mathbf{13}]\big),\\
\mathcal M(1,\tfrac12)\propto\big(\langle\mathbf{13}\rangle+[\mathbf{13}]\big),
\end{array}
\end{equation}
and for $\phi\to ff$, $s_3=0$ with $s_1=s_2=\tfrac12$, the two combinations $(0,0)$ and $(1,1)$ give
\begin{equation}\label{eq:sm-Sff}
\begin{array}{@{}l@{}}
\mathcal M(0,0)\propto\big(\langle\mathbf{12}\rangle-[\mathbf{12}]\big),\\
\mathcal M(1,1)\propto\big(\langle\mathbf{12}\rangle+[\mathbf{12}]\big),
\end{array}
\end{equation}
for $f\to fV$, $s_3=\tfrac12$ with $s_1=\tfrac12$ and $s_2=1$, the four combinations $(0,\tfrac12)$, $(1,\tfrac12)$, $(1,\tfrac32)$ and $(2,\tfrac32)$ give
\begin{equation}\label{eq:sm-FfV}
\begin{array}{@{}l@{}}
\mathcal M(0,\tfrac12)\propto\big(\langle\mathbf{12}\rangle-[\mathbf{12}]\big)\big(\langle\mathbf{23}\rangle-[\mathbf{23}]\big),\\
\mathcal M(1,\tfrac12)\propto\big(\langle\mathbf{12}\rangle-[\mathbf{12}]\big)\big(\langle\mathbf{23}\rangle+[\mathbf{23}]\big),\\
\mathcal M(1,\tfrac32)\propto(m_2-m_3-m_1)\big(\langle\mathbf{12}\rangle+[\mathbf{12}]\big)\big(\langle\mathbf{23}\rangle-[\mathbf{23}]\big)\\
\qquad+\langle\mathbf{2}|p_1|\mathbf{2}]\big(\langle\mathbf{13}\rangle-[\mathbf{13}]\big),\\
\mathcal M(2,\tfrac32)\propto(m_3+m_2-m_1)\big(\langle\mathbf{12}\rangle+[\mathbf{12}]\big)\big(\langle\mathbf{23}\rangle+[\mathbf{23}]\big)\\
\qquad+\langle\mathbf{2}|1|\mathbf{2}]\big(\langle\mathbf{13}\rangle+[\mathbf{13}]\big),
\end{array}
\end{equation}
for $V\to ff$, $s_3=1$ with $s_1=s_2=\tfrac12$, the four combinations $(1,0)$, $(0,1)$, $(1,1)$ and $(2,1)$ give
\begin{equation}\label{eq:sm-Vff}
\begin{array}{@{}l@{}}
\mathcal M(1,0)\propto\big(\langle\mathbf{12}\rangle-[\mathbf{12}]\big)\,\langle\mathbf{3}|p_1|\mathbf{3}],\\
\mathcal M(0,1)\propto\big(\langle\mathbf{13}\rangle-[\mathbf{13}]\big)\big(\langle\mathbf{23}\rangle-[\mathbf{23}]\big),\\
\mathcal M(1,1)\propto(m_3+m_1-m_2)\big(\langle\mathbf{13}\rangle+[\mathbf{13}]\big)\big(\langle\mathbf{23}\rangle-[\mathbf{23}]\big)\\
\qquad-(m_3+m_2-m_1)\big(\langle\mathbf{13}\rangle-[\mathbf{13}]\big)\big(\langle\mathbf{23}\rangle+[\mathbf{23}]\big),\\
\mathcal M(2,1)\propto\big(\langle\mathbf{12}\rangle+[\mathbf{12}]\big)\,\langle\mathbf{3}|p_1|\mathbf{3}]\\
\qquad+\tfrac{m_1+m_2+m_3}{4}\big(\langle\mathbf{13}\rangle+[\mathbf{13}]\big)\big(\langle\mathbf{23}\rangle+[\mathbf{23}]\big),
\end{array}
\end{equation}
for $V\to\phi\phi$, $s_3=1$ with $s_1=s_2=0$, the single combination $(1,0)$,
\begin{equation}\label{eq:sm-VSS}
\mathcal M(1,0)\propto\langle\mathbf{3}|p_1|\mathbf{3}],
\end{equation}
for $\phi\to V\phi$, $s_3=0$ with $s_1=1$ and $s_2=0$, the single combination $(1,1)$,
\begin{equation}\label{eq:sm-SVS}
\mathcal M(1,1)\propto\langle\mathbf{1}|p_3|\mathbf{1}],
\end{equation}
for $V\to V\phi$, $s_3=1$ with $s_1=1$ and $s_2=0$, the three combinations $(0,1)$, $(1,1)$ and $(2,1)$,
\begin{equation}\label{eq:sm-VVS2}
\begin{array}{@{}l@{}}
\mathcal M(0,1)\propto\big(\langle\mathbf{13}\rangle-[\mathbf{13}]\big)^2,\\
\mathcal M(1,1)\propto\langle\mathbf{13}\rangle^2-[\mathbf{13}]^2,\\
\mathcal M(2,1)\propto\tfrac{(m_1+m_2+m_3)(m_3+m_1-m_2)}{4}\big(\langle\mathbf{13}\rangle+[\mathbf{13}]\big)^2\\
\qquad-\langle\mathbf{1}|p_3|\mathbf{1}]\,\langle\mathbf{3}|p_1|\mathbf{3}],
\end{array}
\end{equation}
and for $\phi\to VV$, $s_3=0$ with $s_1=s_2=1$, the three combinations $(0,0)$, $(1,1)$ and $(2,2)$,
\begin{equation}\label{eq:sm-SVV}
\begin{array}{@{}l@{}}
\mathcal M(0,0)\propto\big(\langle\mathbf{12}\rangle-[\mathbf{12}]\big)^2,\\
\mathcal M(1,1)\propto\langle\mathbf{12}\rangle^2-[\mathbf{12}]^2,\\
\mathcal M(2,2)\propto\big((m_1-m_2)^2-m_3^2\big)\big(\langle\mathbf{12}\rangle+[\mathbf{12}]\big)^2\\
\qquad+\langle\mathbf{1}|p_3|\mathbf{1}]\,\langle\mathbf{2}|p_3|\mathbf{2}],
\end{array}
\end{equation}

For $V\to VV$ there are seven partial waves in total, as follows
{\setlength{\abovedisplayskip}{3pt plus 1pt minus 1pt}
\setlength{\belowdisplayskip}{3pt plus 1pt minus 1pt}
\setlength{\abovedisplayshortskip}{2pt plus 1pt}
\setlength{\belowdisplayshortskip}{2pt plus 1pt}
\begin{subequations}\label{eq:sm-VVVseven}
\begin{equation}\label{eq:sm-VVVseven1}
\begin{split}
\mathcal M(1,0)&\propto(\mathbf{12})_-^2\,\langle\mathbf{3}|p_1|\mathbf{3}],
\end{split}
\end{equation}
\begin{equation}\label{eq:sm-VVVseven2}
\begin{split}
\mathcal M(0,1)&\propto(\mathbf{12})_-(\mathbf{13})_-(\mathbf{23})_-,
\end{split}
\end{equation}
\begin{equation}\label{eq:sm-VVVseven3}
\begin{split}
\mathcal M(1,1)&\propto(\mathbf{12})_-\Big[(m_3+m_1-m_2)(\mathbf{13})_+(\mathbf{23})_-\\
&\qquad-(m_3+m_2-m_1)(\mathbf{13})_-(\mathbf{23})_+\Big],
\end{split}
\end{equation}
\begin{equation}\label{eq:sm-VVVseven4}
\begin{split}
\mathcal M(2,1)&\propto\tfrac{m_1+m_2+m_3}{4}(\mathbf{12})_-(\mathbf{13})_+(\mathbf{23})_+\\
&\quad+(\mathbf{12})_-(\mathbf{12})_+\,\langle\mathbf{3}|p_1|\mathbf{3}],
\end{split}
\end{equation}
\begin{equation}\label{eq:sm-VVVseven5}
\begin{split}
\mathcal M(1,2)&\propto\big((m_1-m_2)^2-m_3^2\big)(\mathbf{13})_-(\mathbf{23})_-(\mathbf{12})_+\\
&\quad+(m_2-m_3-m_1)\langle\mathbf{1}|p_3|\mathbf{1}](\mathbf{23})_-^2\\
&\quad+(m_3+m_2-m_1)\langle\mathbf{2}|p_3|\mathbf{2}](\mathbf{13})_-^2,
\end{split}
\end{equation}
\begin{equation}\label{eq:sm-VVVseven6}
\begin{split}
\mathcal M(2,2)&\propto(\mathbf{12})_+\Big[(m_2-m_3-m_1)(\mathbf{13})_+(\mathbf{23})_-\\
&\qquad+(m_3+m_2-m_1)(\mathbf{13})_-(\mathbf{23})_+\Big]\\
&\quad+\langle\mathbf{1}|p_3|\mathbf{1}](\mathbf{23})_-^2+\langle\mathbf{2}|p_3|\mathbf{2}](\mathbf{13})_-^2,
\end{split}
\end{equation}
\begin{equation}\label{eq:sm-VVVseven7}
\begin{split}
\mathcal M(3,2)&\propto\Big[\big(m_3^2-(m_1-m_2)^2\big)(\mathbf{12})_+^2\\
&\qquad-\langle\mathbf{1}|p_3|\mathbf{1}]\,\langle\mathbf{2}|p_3|\mathbf{3}]\Big]\langle\mathbf{3}|p_1|\mathbf{3}]\\
&\quad+\tfrac{m_1+m_2+m_3}{4}\Big[(m_1-m_3-m_2)(\mathbf{23})_+^2\langle\mathbf{1}|p_3|\mathbf{1}]\\
&\qquad+(m_3+m_1-m_2)(\mathbf{13})_+^2\langle\mathbf{2}|p_3|\mathbf{2}]\\
&\qquad+\big(m_3^2-(m_1-m_2)^2\big)(\mathbf{12})_+(\mathbf{13})_+(\mathbf{23})_+\Big],
\end{split}
\end{equation}
\end{subequations}
where $(\mathbf{ij})_{\pm}\equiv\langle\mathbf{ij}\rangle\pm[\mathbf{ij}]$, each line being a polynomial in $(\mathbf{ij})_{\pm}$ and in $\langle\mathbf{i}|p_j|\mathbf{i}]$. These seven structures are equivalent to the seven collected in Table~\ref{tab:3ptstructures}.

\medskip
\textbf{\textit{From $J^{PC}$ to $LS$ in bilinear operators.}} The $LS$ analysis above is not only a way of constructing amplitudes: the same structure serves to analyse form factors, since the matrix element of an operator can be resolved in the same basis. 

\textit{Form factor of fermion bilinear operators.} Consider, for instance, the annihilation of two spinning particles,
\begin{equation}\label{eq:formfactor}
\langle 0|\mathcal O_i|1,2\rangle,
\end{equation}
where $\mathcal O_i$ is a bilinear local operator, in general not a Lorentz scalar. For instance, it may be one of
\begin{equation}\label{eq:bilinears}
\bar\psi\psi,\quad i\bar\psi\gamma_5\psi,\quad \bar\psi\gamma_0\psi,\quad \bar\psi\gamma_i\psi ,
\end{equation}
which we use as the examples below. This matrix element can be decomposed by $LS$ analysis; we use the spinor formalism to assign definite $LS$ quantum numbers. For a two-particle state the discrete quantum numbers are
\begin{equation}
C=(-1)^{L+S},\qquad P=(-1)^{L+2s},
\end{equation}
with $C$ and $P$ the charge-conjugation and parity eigenvalues. The quantum numbers of a given operator are collected in the triple $(L,S,J^{PC})$, with $J$ the total angular momentum obtained by coupling $L$ and $S$, and $C$, $P$ the eigenvalues above.

The form factor in Eq.~\eqref{eq:formfactor} corresponds, in the COM frame, to a three-point amplitude in which the operator insertion plays the role of the third particle, and in the covariant spinor language this amplitude can be written in an arbitrary frame. Writing the operators above in terms of their time and space components corresponds to the decomposition at rest, $p=(m,\mathbf 0)$: in that frame the zeroth component of a Lorentz tensor is obtained by contracting it with $p_3/m_3$, and its spatial components by contracting it with the polarization vectors $\epsilon_i$. The operators $\bar\psi\gamma_0\psi$ and $\bar\psi\gamma_i\psi$, the third and fourth in Eq.~\eqref{eq:bilinears}, are components of the bilinear current $j^\mu\equiv\bar\psi\gamma_\mu\psi$: the former is the contraction $p_{3\mu}j^\mu/m_3$ and the latter the contraction $\epsilon_{i\mu}j^\mu$. Placed in the three-point amplitude, the operator insertions therefore read

\textit{The spinor form of the bilinears.} In the Weyl basis the fermion field of a leg decomposes as
\begin{equation}\label{eq:chiraldecomp}
\psi\to\binom{|\mathbf 2\rangle}{|\mathbf 2]},\qquad
\bar\psi\to\big([\mathbf 1|,\ \langle\mathbf 1|\big),
\end{equation}
so that every bilinear below reduces to the spinor contractions used in this paper. With this decomposition the spinor representations are
\begin{itemize}
\item $\bar\psi\psi\to[\mathbf{12}]+\langle\mathbf{12}\rangle\in(1,1,0^{++})$;
\item $i\bar\psi\gamma_5\psi\to[\mathbf{12}]-\langle\mathbf{12}\rangle\in(0,0,0^{-+})$;
\item $\bar\psi\gamma_0\psi\to p_3^\mu\bar\psi\gamma_\mu\psi\to[\mathbf{2}|3|\mathbf{1}\rangle+\langle\mathbf{2}|3|\mathbf{1}]=0$;
\item $\bar\psi\gamma_i\psi\to\epsilon_3^\mu\bar\psi\gamma_\mu\psi\to[\mathbf{23}]\langle\mathbf{31}\rangle+\langle\mathbf{23}\rangle[\mathbf{31}]$.
The first three are matched directly; for the fourth the matching is less obvious. Expanded with Eq.~\eqref{eq:sm-Vff}, it splits into two $LS$ channels: combining the rearrangement
\begin{equation}\label{eq:rearrangement}
\begin{split}
[\mathbf{23}]&\langle\mathbf{31}\rangle+\langle\mathbf{23}\rangle[\mathbf{31}]\\
&=\tfrac12\Big((\mathbf{31})_-(\mathbf{32})_--(\mathbf{31})_+(\mathbf{32})_+\Big),
\end{split}
\end{equation}
with the identity
\begin{equation}\label{eq:schouten}
\begin{split}
(\mathbf{12})_+&[\mathbf{3}|p_1|\mathbf{3}\rangle=\tfrac{m_3}{2}\Big((\mathbf{13})_+(\mathbf{23})_++(\mathbf{13})_-(\mathbf{23})_-\Big)\\
&\quad+\tfrac{m_1+m_2}{2}\Big((\mathbf{13})_+(\mathbf{23})_+-(\mathbf{13})_-(\mathbf{23})_-\Big),
\end{split}
\end{equation}
obtained by using the Schouten identity on the first term of $\mathcal A(2,1)$ in Eq.~\eqref{eq:sm-Vff}, one immediately obtains the two channels $\bar\psi\gamma_i\psi\in(2,1,1^{--})$ and $\bar\psi\gamma_i\psi\in(0,1,1^{--})$.
\end{itemize}

Each line is obtained in the same steps: the operator is contracted with $p_3$ or with $\epsilon_3$, the fields $\psi$ and $\bar\psi$ are replaced by the wavefunctions in the spinor-helicity form, and the result is reduced with the spinor techniques. This procedure applies to the remaining fermion bilinears below, where $\sigma^{\mu\nu}\equiv\frac{i}{2}[\gamma^\mu,\gamma^\nu]$:
\begin{gather}
\bar\psi\gamma_5\gamma_0\psi\to p_3^\mu\,\bar\psi\gamma_5\gamma_\mu\psi\to[\boldsymbol{21}]-\langle\boldsymbol{21}\rangle\to(0,0,0^{-+}),\\
\bar\psi\gamma_5\gamma_i\psi\to \epsilon_3^\mu\,\bar\psi\gamma_5\gamma_\mu\psi\to[\boldsymbol{23}]\langle\boldsymbol{31}\rangle-\langle\boldsymbol{23}\rangle[\boldsymbol{31}]\to(1,1,1^{++}),\\
\bar\psi\sigma^{0i}\psi\to p_3^\mu\epsilon_3^\nu\,\bar\psi\sigma_{\mu\nu}\psi\to[\boldsymbol{23}][\boldsymbol{31}]+\langle\boldsymbol{23}\rangle\langle\boldsymbol{31}\rangle\nonumber\\\to(0,1,1^{--})+(2,1,1^{--}),\\
\bar\psi\sigma^{ij}\psi\to \epsilon_3^\mu\epsilon_3^\nu\,\bar\psi\sigma_{\mu\nu}\psi\to[\boldsymbol{23}][\boldsymbol{31}]-\langle\boldsymbol{23}\rangle\langle\boldsymbol{31}\rangle\to(1,0,1^{+-}),
\end{gather}
For the first line the contraction $[\mathbf{2}|3|\mathbf{1}\rangle-\langle\mathbf{2}|3|\mathbf{1}]$ reduces, after integration by parts and use of the equations of motion, to $[\boldsymbol{21}]-\langle\boldsymbol{21}\rangle$. The second line holds for $m_1=m_2$, the third uses the same technique as for $\bar\psi\gamma_i\psi$, and the fourth follows from $\langle 3^I3^J\rangle=-m_3\epsilon^{IJ}$ and $[3^I3^J]=m_3 \epsilon^{IJ}$.

\textit{From spinor-helicity form to non-relativistic multipole expansion.} By the same procedure each of these fermion bilinears can be expanded in the C$LS$ amplitudes; in the COM frame the C$LS$ amplitude reads
\begin{equation}
\mathcal A(L,S)_{\{K\}}^{\{I\}\{J\}}\propto C^{s_1\{I\};s_2\{J\}}_{S\{N\}}\,C^{L\{B\};S\{N\}}_{s_3\{K\}}\,
    [\mathbf{3}|1|\mathbf{3}\rangle^L_{\{B\}},
\end{equation}
where $I$, $J$, $K$ denote the little-group indices. The Clebsch--Gordan coefficients are fixed by the little-group algebra: for $s_1=s_2=\tfrac12$ the spin-coupling ones are
\begin{equation}
C^{\frac{1}{2}I_1;\frac{1}{2}I_2}_{0}\propto\epsilon^{I_1I_2} ,\qquad
C^{\frac{1}{2}I_1;\frac{1}{2}I_2}_{1 J_1 J_2}\propto\delta^{I_1}_{(J_1}\delta^{I_2}_{J_2)},
\end{equation}
and the $LS$ ones are
\begin{gather}
C^{0;1I_1I_2}_{1 J_1 J_2}=C^{1I_1I_2;0}_{1 J_1 J_2}\propto\delta^{(I_1}_{(J_1}\delta^{I_2)}_{J_2)},\\
C^{1;I_1I_2;1K_1K_2}_{1 J_1 J_2}\propto
\delta_{(J_1}^{(I_1}\epsilon^{I_2)(K_1}\delta^{K_2)}_{J_2)},\\
C^{2;I_1I_2I_3I_4;1K_1K_2}_{1 J_1 J_2}\propto\delta_{J_1}^{I_1}\delta^{I_2}_{J_2}\epsilon^{I_3K_1}\epsilon^{I_4K_2}+\text{perm},
\end{gather}
with all index sets $\{I\}$, $\{J\}$, $\{K\}$ totally symmetric. The orbital part is carried by the harmonic tensors
\begin{equation}
Y_0=1,\qquad Y_1=[\mathbf{3}|p_1|\mathbf{3}\rangle,\qquad Y_2=[\mathbf{3}|p_1|\mathbf{3}\rangle^2 ,
\end{equation}
the same objects may also be written with their little-group indices; in the COM frame they read
\begin{equation}
Y_1^{I_1I_2}=\epsilon^{I_1K}(\hat{\mathbf n}\cdot\sigma)_K{}^{I_2},\qquad
Y_2^{\{I\}}=Y_1^{(I_1I_2}Y_1^{I_3I_4)} .
\end{equation}
Here $(\sigma_i)_K{}^J$ are the Pauli matrices acting on the little-group indices, to be distinguished from the spin label $\sigma$. For instance, for $(L,S,J)=(0,0,0)$ and $(1,1,0)$ this gives, respectively,
\begin{equation}
\mathcal A^{I_1 I_2}(0,0)\propto \epsilon^{I_1 I_2},\qquad
\mathcal A^{I_1 I_2}(1,1)\propto \epsilon^{I_1 K}(\hat{\mathbf n}\cdot \sigma)_{K}{}^{I_2} .
\end{equation}
For the part with $J=1$ we use a vector index: writing the amplitude as $\mathcal A_{J_1J_2}^{I_1I_2}$, we represent it by
\begin{equation}
\mathcal A_i^{I_1I_2}\propto\epsilon^{J_1K}(\sigma_i)_K{}^{J_2}\,\mathcal A_{J_1J_2}^{I_1I_2} .
\end{equation}
As a check, the $(1,0)$ structure reproduces the vector form,
\begin{equation}
\begin{split}
\mathcal A_i^{I_1I_2}(1,0)
&=\epsilon^{J_1K}(\sigma_i)_K{}^{J_2}\,\mathcal A_{J_1J_2}^{I_1I_2}\\
&\propto\epsilon^{J_1K}(\sigma_i)_K{}^{J_2}\,
 \epsilon^{I_1I_2}\epsilon^{J_1'K'}(\hat{\mathbf n}\cdot\sigma)_{K'}{}^{J_2'}\,
 \epsilon_{J_1'J_1}\epsilon_{J_2'J_2}\\
&=\epsilon^{I_1I_2}\,
 \epsilon^{J_1K}(\sigma_i)_K{}^{J_2}(\hat{\mathbf n}\cdot\sigma)_{J_1}{}^{J_2'}\,
 \epsilon_{J_2'J_2}\\
&=2\,\epsilon^{I_1I_2}\hat{n}_i ,
\end{split}
\end{equation}
where the last step uses the Pauli completeness relation in the trace form
\begin{equation}\label{eq:sm-paulicomplete}
\mathrm{Tr}\big(\sigma_2\,\sigma_i^{T}\sigma_2\,\sigma_j\big)=2\,\delta_{ij},
\end{equation}
written with $\epsilon^{IJ}=\epsilon_{IJ}=(-i\sigma_2)^{IJ}$, which turns each $\epsilon$ contracted in the second line into a factor of $i\sigma_2$.
With this notation the four structures read
\begin{gather}
\mathcal A_i^{I_1I_2}(1,0)\propto\epsilon^{I_1I_2}\hat{n}_i,\\
\mathcal A_i^{I_1I_2}(0,1)\propto\epsilon^{I_1K}(\sigma_i)_K{}^{I_2},\\
\mathcal A_i^{I_1I_2}(1,1)\propto\epsilon^{I_1K}(\sigma\times\hat{n})_K{}^{I_2},\\
\mathcal A_i^{I_1I_2}(2,1)\propto\epsilon^{I_1K}\bigl(3\hat{n}_i\,\hat{n}\cdot\sigma-\sigma_i\bigr)_K{}^{I_2}.
\end{gather}
The arguments label the the orbital and spin angular momenta of each structure.

\medskip
\textbf{\textit{General form factors and correlation functions.}} The construction of the previous section generalises to a general form factor structure and to a general frame. 
\textit{Form factor construction.} For the general form factor
\begin{equation}\label{eq:sm-defF}
F^{\{\alpha\}\{\dot\alpha\}}(p)=\int d^4x\,e^{ip\cdot x}\,
\langle 0|\mathcal O^{\{\alpha\}\{\dot\alpha\}}(x)|p_1,\sigma_1,p_2,\sigma_2\rangle ,
\end{equation}
with $p=p_1+p_2$, arbitrary external states $|p_1,\sigma_1\rangle$ and $|p_2,\sigma_2\rangle$ of spins $s_1$ and $s_2$, and the operator $\mathcal O$ carrying the spinor indices of the Lorentz representation $(j_L,j_R)$, all independent and complete Lorentz structures can be built from spinors without returning to the COM frame. It suffices to define the object that carries little-group rather than Lorentz indices, the index sets $\{I\}$ and $\{J\}$ being of rank $2j_L$ and $2j_R$,
\begin{equation}\label{eq:sm-tildeF}
\tilde F_{\{I\} \{J\}}\equiv F^{\{\alpha\}\{\dot\alpha\}}\,
\lambda^{2j_L}_{\{\alpha\}\{I\}}\,\tilde\lambda^{2j_R}_{\{\dot\alpha\}\{J\}},
\end{equation}
Its inverse follows by contracting with the dual tensors,
\begin{equation}\label{eq:sm-Finv}
F^{\{\alpha\}\{\dot\alpha\}}
=\frac{1}{m^{2j_L+2j_R}}\tilde F_{\{I\}\{J\}}\,
\big(\lambda^{2j_L}\big)^{\{\alpha\}\{I\}}
\big(\tilde\lambda^{2j_R}\big)^{\{\dot\alpha\}\{J\}},
\end{equation}
the structures of $\tilde F$ being in one-to-one correspondence with those of $F$. Decomposing $\tilde F$ into $SU(2)$ irreducible parts,
\begin{equation}\label{eq:sm-Fj}
\tilde F_{j\{K\}}\equiv C^{j_L \{I\};j_R \{J\}}_{j\{K\}}\tilde F_{\{I\} \{J\}},
\end{equation}
\begin{equation}\label{eq:sm-Fjinv}
  \tilde F_{\{I\} \{J\}}=\sum_j C_{j_L \{I\};j_R \{J\}}^{j\{K\}} \tilde F_{j\{K\}}
\end{equation}
with $j=|j_L-j_R|,\ldots,j_L+j_R$, the structures of $\tilde F_j$ are exactly those of the amplitude for $j\to s_1+s_2$, given by Eq.~\eqref{eq:final3ptampLS} and expanded as
\begin{equation}\label{eq:sm-Fjexp}
\tilde F_j=\sum_{L,S}g_{jLS}\,\mathcal M(L,S,j).
\end{equation}
Substituting Eq.~\eqref{eq:sm-Fjexp} into the inverse relations \eqref{eq:sm-Fjinv} and \eqref{eq:sm-Finv} yields the complete set of Lorentz structures of $F$,
\begin{equation}\label{eq:sm-Fstructures}
\begin{split}
F^{\{\alpha\}\{\dot\alpha\}}&=\frac{1}{m^{2j_L+2j_R}}\sum_{j,L,S}g_{jLS}\,
\mathcal M(L,S,j)_{\{K\}}\,
\\
&\times C^{j\{K\}}_{j_L\{I\};j_R\{J\}}\big(\lambda^{2j_L}\big)^{\{\alpha\}\{I\}}
\big(\tilde\lambda^{2j_R}\big)^{\{\dot\alpha\}\{J\}} .
\end{split}
\end{equation}
The structures are therefore built from spinor variables, and their number follows from the same spin-coupling technique as that of the amplitudes~\cite{Huang:2026kfq}.

\textit{Parton correlation functions.} These are the special case of a form factor structure in which the two external momenta coincide and the relative momentum $q$ is carried by the internal operator,
\begin{equation*}
C^{\{\alpha\}\{\dot\alpha\}}(p,q)=\int dx\,\rho_{\sigma\sigma'}\langle p\sigma|e^{iq\cdot x}\mathcal{O}^{\{\alpha\}\{\dot\alpha\}}(x,0)|p\sigma'\rangle,
\end{equation*}
with $\rho$ the density matrix. Contracting with the spinors reduces the problem to a little-group tensor, $C^{\{\alpha\}\{\dot\alpha\}}(\lambda^{2j_l})^{\{I\}}_{\{\alpha\}}(\tilde\lambda^{2j_r})^{\{J\}}_{\{\dot\alpha\}}=\rho_{\sigma\sigma'}\tilde C^{\{I\}\{J\}}_{\sigma\sigma'}$, whose independent forms follow from decomposing the direct product of all little-group indices into a scalar structure. The density matrix then serves as a general coefficient, and the approach reproduces the known quark--quark and gluon--gluon correlations.

\medskip
\textbf{\textit{Covariant partial wave analysis on cascade decay.}} Traditionally the partial-wave decomposition is performed through the coupling of the orbital and spin parts in the two-body COM frame; when the process involves cascade decay chains the momenta must be boosted and the resulting amplitudes are not covariant, the orbital and spin parts being mixed~\cite{Habermann:2024sxs}. With canonical particle states the $LS$ decomposition can be performed directly among particles in an arbitrary frame, without boost transformations; this partial-wave analysis~\cite{Chung:1993da,Chung:1997jn,Zou:2002ar} has been applied to the process $\Lambda_c\to\Lambda\pi^+\pi^0$~\cite{Huang:2026egv}.

This cascade decay contains several possible intermediate resonant states, with each resonance corresponding to a distinct decay chain. We consider the intermediate states that Ref.~\cite{BESIII:2022udq} reports above $5\sigma$, namely the $\rho(770)^{+}$, $\Sigma(1385)^{+}$, $\Sigma(1385)^{0}$, $\Sigma(1670)^{+}$, $\Sigma(1670)^{0}$, $\Sigma(1750)^{+}$ and $\Sigma(1750)^{0}$ states together with the nonresonant $NR_{1^{-}}$. These components fall into three topological classes:
\begin{itemize}
	\item \textbf{Topology I} ($\rho^+$-type): $\Lambda_c^+ \to \Lambda R$, $R \to \pi^+ \pi^0$, $\Lambda \to p \pi^-$,
	with meson resonance $R$ decaying in $\pi^+\pi^0$; the two leading terms are the \textbf{$\rho(770)^+$} and the nonresonant $J^P=1^-$ one, written $NR_{1^-}$. The amplitudes corresponding to these channel are defined as:
	\begin{equation}\label{eq:sm-rho-canonical}
\mathcal{A}_{\Lambda^+_c}^{p\pi^-\pi^+\pi^0}(\rho) = \mathcal{A}_{\Lambda^+_c}^{\Lambda\rho}\mathcal{A}_{\rho}^{\pi^+\pi^0}\mathcal{A}_{\Lambda}^{p\pi^-},
\end{equation}
	\begin{equation}\label{eq:sm-NR-canonical}
\mathcal{A}_{\Lambda^+_c}^{p\pi^-\pi^+\pi^0}(NR_{1^-}) =\mathcal{A}_{\Lambda^+_c}^{\Lambda NR_{1^-}}\mathcal{A}_{NR_{1^-}}^{\pi^+\pi^0}\mathcal{A}_{\Lambda}^{p\pi^-},
\end{equation}
	where each particle symbol stands for the particle together with its little-group indices, $p$, $\Lambda_c^+$, $\Lambda$, $\rho$; the scalars $\pi$ carry the trivial little-group index, that is none, and appear only as labels.

	\item \textbf{Topology II} ($\Sigma^{*+}$-type): $\Lambda_c^+ \to \Sigma^{*+} \pi^0$, $\Sigma^{*+} \to \Lambda \pi^+$, $\Lambda \to p \pi^-$,
	with $\Sigma^{*+}$ an excited baryon; the candidates are the \textbf{$\Sigma(1385)^+$}, \textbf{$\Sigma(1670)^+$} and \textbf{$\Sigma(1750)^+$}, and
	\begin{equation}\label{eq:sm-sigmaplus-canonical}
\mathcal{A}_{\Lambda^+_c}^{p\pi^-\pi^+\pi^0}(\Sigma^{*+}) = \mathcal{A}_{\Lambda^+_c}^{\Sigma^{*+}\pi^0}\mathcal{A}_{\Sigma^{*+}}^{\Lambda\pi^+}\mathcal{A}_{\Lambda}^{p\pi^-}.
\end{equation}

	\item \textbf{Topology III} ($\Sigma^{*0}$-type): $\Lambda_c^+ \to \Sigma^{*0} \pi^+$, $\Sigma^{*0} \to \Lambda \pi^0$, $\Lambda \to p \pi^-$,
	with $\Sigma^{*0}$ another excited baryon; the candidates are the \textbf{$\Sigma(1385)^0$}, \textbf{$\Sigma(1670)^0$} and \textbf{$\Sigma(1750)^0$}, and
	\begin{equation}\label{eq:sm-sigma0-canonical}
\mathcal{A}_{\Lambda^+_c}^{p\pi^-\pi^+\pi^0}(\Sigma^{*0}) = \mathcal{A}_{\Lambda^+_c}^{\Sigma^{*0}\pi^+}\mathcal{A}_{\Sigma^{*0}}^{\Lambda\pi^0}\mathcal{A}_{\Lambda}^{p\pi^-}.
\end{equation}
\end{itemize}
In each topology every three-point amplitude can be analysed in the $LS$ basis; bootstrapping them gives the total amplitude of that topology, and the amplitude of the whole $\Lambda_c^+$ decay is the sum of the amplitudes of the topologies. Using this amplitude basis together with the open-source package TF-PWA~\cite{tfpwa}, events are generated and fitted; the comparison with the conventional treatments carried out in Ref.~\cite{Huang:2026egv} shows that the results agree very well.

\end{document}